\documentclass[times,twocolumn,ifnopreprintline, review]{elsarticle}
\usepackage{medima}
\usepackage[dvipsnames]{xcolor}
\usepackage{multicol}
\usepackage{soul}

\usepackage{framed,multirow}
\usepackage[utf8]{inputenc}
\usepackage{latexsym}
\usepackage{url}
\usepackage{subfig}
\usepackage{booktabs}
\usepackage{graphicx}
\usepackage{cite}
\usepackage[bookmarks,bookmarksnumbered]{hyperref}

\usepackage[fleqn]{amsmath}
\usepackage{amssymb,amsfonts}
\usepackage{algorithmic}
\usepackage{textcomp}
\usepackage{bbding}
\usepackage{mathtools}
\usepackage{pifont}% http://ctan.org/pkg/pifont
\usepackage[normalem]{ulem}
\usepackage{algorithmic,algorithm}
\usepackage{mathtools}
\usepackage{enumitem}
\useunder{\uline}{\ul}{}
\usepackage{threeparttable}
\hypersetup{colorlinks = green,linkcolor = green,anchorcolor =green,citecolor = green,filecolor = green,urlcolor = green,pdfauthor=author,citecolor=green}
\newcommand{\revise}[1]{\textcolor{black}{#1}}
\newcommand{\secondrevise}[1]{\textcolor{black}{#1}}
\setstcolor{red}

\newcommand{\randchain}{\textit{RandChain}}

\newcommand{\advchain}{\textit{AdvChain}}

\DeclareMathOperator{\EX}{\mathbb{E}}% expected value

\def\BibTeX{{\rm B\kern-.05em{\sc i\kern-.025em b}\kern-.08em
    T\kern-.1667em\lower.7ex\hbox{E}\kern-.125emX}}
\newcommand{\crossmark}{\ding{55}}%

\newcommand{\Reals}{\mathbb{R}}

\newcommand{\classifier}{\mathit{f}}

\newcommand{\prediction}{\mathbf{p}}
\newcommand{\p}{\mathbf{p}}

\newcommand{\image}{\mathbf{x}}

\newcommand{\imagespace}{\mathcal{X}}
\newcommand{\segmentation}{\mathbf{y}}
\newcommand{\labelspace}{\mathcal{Y}}
\newcommand{\dataset}{D}

\newcommand{\condition}{\mathcal{C}}

\newcommand{\x}{\mathbf{x}}

\newcommand{\xpos}{u}
\newcommand{\ypos}{v}

\newcommand{\0}{\mathbf{0}}

\newcommand{\transformation}{\mathbf{t}}

\newcommand{\noise}{\mathbf{r}}
\newcommand{\velocity}{\mathbf{v}}
\newcommand{\cpoints}{\mathbf{c}}

\newcommand{\one}{\mathbf{1}}

\newcommand{\augment}{\mathcal{T}}% Caligraphic T for example
\newcommand{\biasfunc}{\augment_{\rm bias}(\image; \cpoints)}

\newcommand{\difffunc}{\augment_{\rm morph}(\image; \velocity)}

\DeclareMathOperator{\proj}{\Pi} 

\newcommand{\Loss}{\mathcal{L}}

\newcommand{\RegLoss}{\mathcal{R}}

\newcommand{\Distance}{\mathcal{D}}

\newcommand{\argmax}{\mathop{\rm argmax}\limits}

\renewcommand{\revise}[1]{\textcolor{black}{#1}}
\usepackage[modulo]{lineno} % FOR DOUBLE COLUMNS
\definecolor{newcolor}{rgb}{.8,.349,.1}

\journal{Medical Image Analysis}

\let\oldequation\equation
\let\oldendequation\endequation
\renewenvironment{equation}
  {\linenomathNonumbers\oldequation}
  {\oldendequation\endlinenomath}

\begin{document}
\verso{Chen Chen \textit{et~al.}}
\begin{frontmatter}
% \title{Type the title of your paper, only capitalize first
% word and proper nouns\tnoteref{tnote1}}%
% \tnotetext[tnote1]{This is an example for title footnote coding.}
\title{Enhancing MR Image Segmentation with Realistic Adversarial Data Augmentation}
\author[1]{Chen \snm{Chen}\corref{cor1}}
\cortext[cor1]{Corresponding author: Chen Chen, 
chen.chen15@imperial.ac.uk
}
% \ead{chen.chen15@imperial.ac.uk}
\author[2]{Chen \snm{Qin}}
\author[1]{Cheng \snm{Ouyang}}
\author[1]{Zeju \snm{Li}}

%% Third author's email
\author[3,4]{Shuo \snm{Wang}}
\author[1]{Huaqi \snm{Qiu}}
\author[1]{Liang \snm{Chen}}
\author[1,5]{Giacomo \snm{Tarroni}}
\author[1,6,7]{Wenjia \snm{Bai}}
\author[1,8]{Daniel \snm{Rueckert}}

\address[1]{Department of Computing, Imperial College London, UK}
\address[2]{Institute for Digital Communications, School of Engineering, University of Edinburgh, UK}
\address[3]{Digital Medicine Research Centre, School of Basic Medical Sciences, Fudan University, China}
\address[4]{Shanghai Key Laboratory of MICCAI, Shanghai, China}
\address[5]{CitAI Research Centre, Department of Computer Science, City, University of London, UK}
\address[6]{Department of Brain Sciences, Imperial College London, UK}
\address[7]{Data Science Institute, Imperial College London, UK}
\address[8]{Klinikum rechts der Isar, Technical University of Munich, Germany}

\received{xx xx 2021}
\finalform{xx xx 2022}
\accepted{xx xx 2022}
\availableonline{xx xx 2022}
\communicated{C. Chen}

\begin{abstract}
The success of neural networks on medical image segmentation tasks typically relies on large labeled datasets for model training. However, acquiring and manually labeling a large medical image set is resource-intensive, expensive, and sometimes impractical due to data sharing and privacy issues. \revise{To address this challenge, we propose \advchain, a generic adversarial data augmentation framework, aiming at improving both the \emph{diversity} and \emph{effectiveness} of training data for medical image segmentation tasks. \advchain\ augments data with \emph{dynamic data augmentation}, generating randomly chained photo-metric and geometric transformations to resemble realistic yet challenging imaging variations to expand training data. By jointly optimizing the data augmentation model and a segmentation network during training, challenging examples are generated to enhance network generalizability for the downstream task. The proposed adversarial data augmentation does not rely on generative networks and can be used as a plug-in module in general segmentation networks.} It is computationally efficient and applicable for both low-shot supervised and semi-supervised learning. We analyze and evaluate the method on two MR image segmentation tasks: cardiac segmentation and prostate segmentation \revise{with limited labeled data}. Results show that the proposed approach can alleviate the need for labeled data while improving model generalization ability, indicating its practical value in medical imaging applications.
\end{abstract}


\end{frontmatter}

\section{Introduction}
%%State the importance of this task
Medical image segmentation plays an essential role in healthcare applications, including disease diagnosis, treatment planning and clinical research~\citep{Smistad_2015_MedIA}.
In recent years, many deep learning-based techniques have been developed for medical image segmentation, achieving high performance in terms of both speed and accuracy~\citep{Shen_2017_Review,litjens_2017_survey}. However, training a deep neural network generally requires a large amount of labeled data. In medical imaging, acquiring and manually labeling such a large dataset is extremely challenging for several reasons. First, labeling medical images is time-consuming and expensive as it requires experienced human observers with domain expertise. Second, collecting and sharing large datasets across clinical sites is difficult due to data privacy and ethical issues. As a result, it is typical that only a small number of labeled images are available for training a neural network, which hinders the deployment of effective deep learning solutions for healthcare applications. 

%% Previous work have addressed
To alleviate the data scarcity problem, data augmentation approaches have been proposed~\citep{Shorten_2019_Big_Data_Survey}, which aim to increase the diversity of the available training data without collecting and manually labeling new data. Conventional data augmentation methods mainly focus on applying simple \emph{random} transformations to labeled images. These random transformations include intensity transformations (e.g. pixel-wise noise or image-wise brightness and contrast adjustment) and geometric transformations (e.g. affine or elastic transformations). Most of these transformations perform basic image manipulations without taking image contents into account or accounting for downstream tasks, which may introduce redundant data points that does not necessarily improve generalization~\citep{Miyato_2018_PAMI_VAT}.  

% Generative network-based methods, which learn to generate samples to match real-world data distributions, by contrast, have demonstrated their capability in capturing and resembling complex image appearances and geometric variations~\citep{Zhao_2019_CVPR_oneshotDA,Chaitanya_2019_IPMI}. However, training generative neural networks is not trivial, which may suffer from training instability and mode collapse problems. Also, it is not computationally efficient as it requires a large amount of training time and computational memory. 

%% Our works.
\revise{
 In this work, we introduce a generic adversarial data augmentation framework (\advchain), aiming at improving both the \emph{diversity} and \emph{effectiveness} of  training data for medical image segmentation tasks. \advchain\ improves data diversity with \emph{dynamic data augmentation}, generating randomly chained photo-metric and geometric transformations to resemble realistic complex data variation at training. Different from conventional random data augmentation approaches, \advchain\ allows to optimize the underlying transformation parameters in arbitrarily chained transformations (providing they are differentiable). By taking both image information and the current network fragility into account to optimize the transformation parameters, \advchain\ improves the `hardness' of augmented images to better regularize the network training (Sec.\ref{sec:3.1})}.
 
\revise{The proposed framework \advchain\ can accommodate a wide range of differentiable photometric and geometric transformations for the joint optimization of data augmentation and network in both supervised and semi-supervised learning. In this work, as a proof of concept, four different image transformation models are employed to resemble realistic imaging variations in MR imaging. They are: a) an image noise augmentation model; b) an intensity transformation model which amplifies intensity non-uniformity by simulating low-frequency intensity corruptions caused by inhomogeneities of the magnetic field; c) a global image geometric transformation model based on affine transformation that simulates patient movement (e.g., rotation, translation) and imaging resolution variations (e.g., scaling) during scanning; d) a diffeomorphic deformation model which simulates intra-subject morphological difference attributed to pathology, growth and motion, and inter-subject morphological difference. By generating realistic and various `hard' examples for data augmentation, we force the network to learn robust semantic features against various imaging variations, leading to improved model generalization. Besides, \advchain\ strengthens the consistency regularization for medical segmentation tasks by employing a composite loss function, which encourages both pixel-level consistency as well as contour-based consistency (Sec.~\ref{SEC:consistency loss}). We demonstrate the efficacy of the proposed method on two public MR image datasets in challenging low-data supervised and semi-supervised settings (e.g., with only 1 labeled subject for training). Our method outperforms several strong consistency-regularized methods and strong composite data augmentation method (RandAugment~\citep{Cubuk_2020_RandAugment}) in low-data regimes (e.g., with only 1 or \secondrevise{3} labeled subject for training), indicating its efficacy to improve the generalisability of the model on MR segmentation tasks when labeled data is limited.}
 
This work is an extension to our previously presented work at MICCAI~\citep{chen_2020_realistic}, where we introduced adversarial photometric data augmentation with a bias field intensity transformation model and demonstrated its effectiveness on a \emph{binary} cardiac segmentation task. In this work, we substantially extend the framework by including both adversarial photometric and geometric transformations and composing these transformations \revise{in a flexible way} to further improve image diversity and resemble data variations in magnetic resonance (MR) imaging. \revise{In particular, we present a novel adversarial diffeomorphic deformation model to generate challenging morphological variations, as a way to improve the segmentation model generalization ability.} Finally, we extend the framework to multi-class segmentation problems and comprehensively evaluate our method on two public datasets, one consisting of cardiac MR images and the other of prostate MR images. Experiments on both datasets show the effectiveness of our method, which improves image segmentation performance and outperforms competitive consistency regularization-based methods.

\section{Related work}
We first review several advanced data augmentation techniques that have been developed recently (e.g. data mixing, adversarial data augmentation) and then introduce consistency-based semi-supervised learning methods which are closely related to this work. 

\subsection{Data mixing} 
Data mixing methods generate new data samples by mixing multiple samples together~\citep{Zhang_2018_ICLR_mixup,HendrycksMCZGL20_ICLR_2020,berthelot_2019_mixmatch}. A representative work is Mixup~\citep{Zhang_2018_ICLR_mixup}, which creates new training samples by combining random pairs of images ($x_a$, $x_b$) and their labels ($y_a$, $y_b$) via linear interpolation: $x^{new}= \beta x_a+(1-\beta)x_b, y^{new}= \beta y_a+(1-\beta)y_b$, where $\beta$ is a weighting parameter sampled from the beta distribution. Though originally proposed for image classification, Mixup has been successfully adapted to medical image segmentation tasks, including knee segmentation~\citep{Panfilov_2019_ICCVW}, brain segmentation~\citep{Li_2019_MICCAI}, and cardiac segmentation~\citep{Chaitanya_2019_IPMI}. One problem with this technique is that the mixed images can be unrealistic and difficult to interpret. Also, the diversity of generated samples by data mixing is limited since the mixed samples still lie in the span of the training data~\citep{Wu_2020_ICML_generalization}.

\subsection{Adversarial data augmentation}
Adversarial data augmentation applies perturbations to original images to fool the model into making classification mistakes. These perturbed images (also known as adversarial images) are then used to optimize the network for improved robustness against particular perturbations. Recent studies have shown that adversarial data augmentation can be more effective than random data augmentation~\citep{Madry_2017_PGDattack,Volpi_2018_NIPS_Generalizing,Suzuki_2020_AAAI_ATSL}. Most existing works are based on simple gradient-based noise attack, i.e. using the gradients of the neural network to generate additive adversarial noise to perturb images~\citep{Madry_2017_PGDattack,Goodfellow_2015_FGSM,Carlini_2017_CW_attack,Tramer_2019_NIPS_Adversarial,Miyato_2018_PAMI_VAT, Paschali_2018_MICCAI}. However, researchers have found that neural networks can be fragile to other more complex forms of transformations that may occur in images, such as affine transformations~\citep{Kanbak_2018_CVPR_Geometric,Engstrom_2019_ICML,Zeng_2019_CVPR_Beyond,Finlayson_2019_Science}, illumination  changes~\citep{Zeng_2019_CVPR_Beyond} or small deformations~\citep{Alaifari_2019_ICLR_ADef}. For medical image segmentation, the majority of related works focus on crafting effective adversarial examples and leverage them to evaluate model robustness. For example, \citet{Paschali_2018_MICCAI} applied a targeted attack, specifically a dense adversary generation (DAG) attack~\citep{Xie_2017_ICCV_DAG}, to generate effective pixel-wise noise, which fools a segmentation network into producing poor segmentation on brain images. Chen et al. \citep{Chen_2019_SASHIMI_Intelligent} proposed to use conditional GANs to model spatial deformation and noises for adversarial image construction. 

\revise{In contrast to existing adversarial data augmentation which augments images with a \emph{single, fixed} type of image transformation~\citep{Miyato_2018_PAMI_VAT,Wang_2021_MedIA_Deep,chen_2020_realistic,Xie_2017_ICCV_DAG,Alaifari_2019_ICLR_ADef,Kanbak_2018_CVPR_Geometric,Zeng_2019_CVPR_Beyond,Finlayson_2019_Science,Madry_2017_PGDattack,Goodfellow_2015_FGSM,Carlini_2017_CW_attack,Tramer_2019_NIPS_Adversarial,Paschali_2018_MICCAI}, \advchain\ is capable of directly optimizing the transformations parameters in \emph{dynamic} data augmentations, e.g., arbitrarily chained image photometric and geometric transformations, better generating realistic and challenging image variations that may occur at medical imaging applications. Existing composite data augmentation optimization works such as generative adversarial network (GAN)-based data augmentation approach~\citep{Gao_2021_IPMI_Enabling_Data_Diversity} are very computational intensive and suffer from the training instability problem, as they need to train different GANs to produce photometric and geometric transformation parameters separately. Their approach can not be used to optimize randomly chained transformations due to high training instability and memory costs, which involves the optimization of multiple stacked GANs. Our method, by contrast, can efficiently optimize all different transformations in a chain, even with only one forward pass and backward pass.} 

\revise{On top of \advchain, we also present a \emph{novel} adversarial data augmentation with diffeomorphic transformations based on stationary velocity fields, which could generate realistic, morphological variations to fool the network. At training, we directly optimize the underlying static velocity field and integrate them to generate diffeomorphic deformations. This is fundamentally different from existing adversarial deformation works based on GANs~\citep{Chen_2019_SASHIMI_Intelligent, Gao_2021_IPMI_Enabling_Data_Diversity} where a generative network is required to model additive displacement fields. The network has to be pre-trained with a carefully designed regularization loss on the deformation fields to restrict the realism of generated deformations. The produced deformations may not be invertible, thus restricting its use for computing the pixel-wise consistency regularization in the original input space. } 

\subsection{Consistency regularization}
Viewing data augmentation as a way of encoding invariances and equivalences into a neural network, consistency regularization methods apply data augmentation to unlabeled data for semi-supervised learning based on the assumption that the predictions of a data point and its augmented/perturbed example should be consistent~\citep{sajjadi_2016_NIPS_regularization,Li_2020_TNNLS_transformation,Miyato_2018_PAMI_VAT,berthelot_2019_mixmatch,Xie_2020_NIPS, sohn_2020_fixmatch,Wang_2021_MedIA_Deep}. A consistency regularization term is generally introduced to the loss function to encourage a model to produce consistent predictions on similar inputs (e.g., unlabeled data and its augmented ones). On the basis of this mechanism, many works explored different data augmentation techniques, including random data augmentation (e.g., pixel-level noise, affine transformations)~\citep{sajjadi_2016_NIPS_regularization,Liu_2020_TMI,Li_2020_TNNLS_transformation}, data mixing~\citep{berthelot_2019_mixmatch,HendrycksMCZGL20_ICLR_2020} and adversarial data augmentation techniques~\citep{Miyato_2018_PAMI_VAT,Volpi_2018_NIPS_Generalizing,Xie_2019_Arxiv,Suzuki_2020_AAAI_ATSL}.

For medical image segmentation tasks, several related works explored different types of data augmentation to enhance consistency regularization~\citep{cui_2019_semi_noise,Li_2020_TNNLS_transformation, chen_2020_realistic}. These works focused on utilizing weak, random augmentation methods such as random Gaussian noise~\citep{cui_2019_semi_noise}, random affine transformations~\citep{Li_2020_TNNLS_transformation}, and adversarial bias fields~\citep{chen_2020_realistic}. A major difference of our work is that we consider modeling more complex photometric and geometric transformations and propose adversarial training to optimize the transformation parameters to generate \emph{more challenging augmented images}. We believe that, with more diverse and effective realistic data augmentation to regularize training, the proposed method can better enforce the model to learn high-level, robust representations for an improved generalization ability. 

% \subsection{Semi-supervised learning in Medical Image Segmentation}
% Semi-supervised learning (SSL), which  leverage both labeled and \emph{unlabeled} data during training, is a promising approach to improve model generalization ability. While there are many different streams for semi-supervised learning (e.g. entropy minimization~\citep{Grandvalet_2005_EntropyMinimization}, generative models~\citep{Kingma_2014_NIPS}, and graph-based methods~\citep{Kipf_2017_ICLR}), in this paper, we mainly review consistency regularization-based methods as they are most relevant to our work. 
% Recent semi-supervised learning methods in medical
% image segmentation mainly can be grouped into:  (1) consistency-based approach, (2) entropy-based approach, (3) GAN-based approach. % Dong et al. [XXX] applies a GAN loss to encourage predicted segmentation maps on unlabeled images to be similar to the ground-truth labels and show that it improves the performance in a semi-supervised setting. Others applies pseudo labeling.

%% TMI： equationref: use (1) not Eq. (1).
\section{Methods}
\label{SEC:methodology}
\begin{figure*}[!ht]
    \centering
    \includegraphics[width=0.95\textwidth]{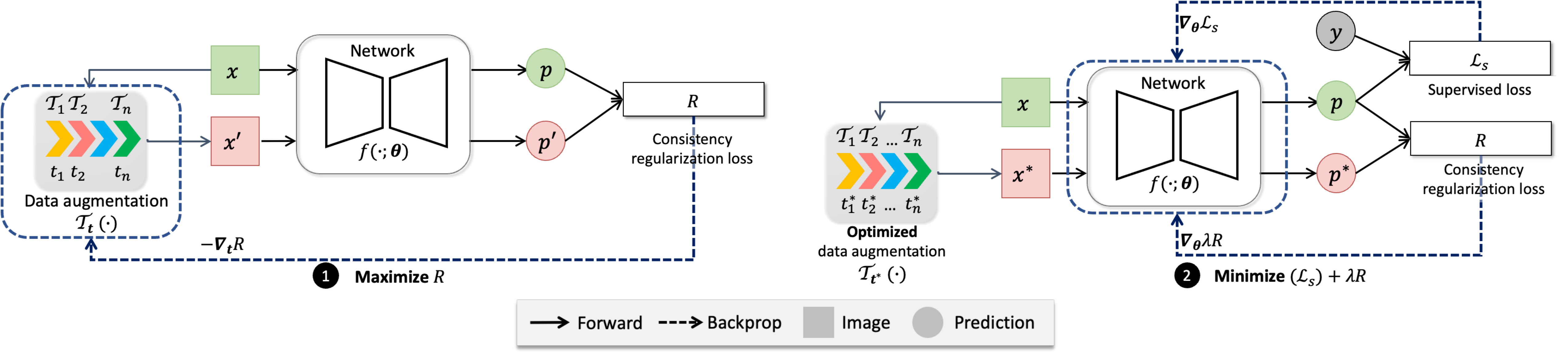}
    \caption{\revise{\textbf{\advchain\ {overview}.} \advchain\ is a generic adversarial data augmentation framework for medical image segmentation, which allows to optimize the parameters in a \emph{randomly sampled} augmentation chain (incl. photo-metric and geometric transformations) for enhanced consistency regularization. \textbf{Left}: Given a segmentation network $\classifier_{\theta}$, an input image $\image$ and a \emph{randomly sampled} chain of transformation functions $\augment: \augment_1 \circ \augment_2,...,\circ \augment_n$ ($n>=1$) with corresponding transformation parameters $\transformation: \transformation_1, \transformation_2,..., \transformation_n$, \advchain\ first optimizes the underlying transformation parameters ${\transformation}$ in the direction of \emph{maximizing} the inconsistency (measured by $\RegLoss$) between the network prediction for the original image $\prediction=\classifier_\theta(\image)$ and the prediction for the augmented image $\prediction'=\classifier_\theta(\image')$. \textbf{Right}: The updated transformation parameters $\transformation^{*}$  are then used to generate an \advchain\ augmented image $\image^*$ to train the network. Specifically, the network parameters $\theta$ are optimized in the direction of \emph{minimizing} the consistency loss $\RegLoss$ computed between the original prediction $\prediction$ and the prediction for the \advchain\ image $\prediction^*$, in together with the supervised loss $\Loss_s$ (if its ground-truth label $\segmentation$ is available). Best viewed in color.}}
    \label{fig:framework}
\end{figure*}
The goal of medical image segmentation is to learn a mapping from an image space $\imagespace$ to a label space $\labelspace$. In deep learning, the mapping is parameterized by a neural network, e.g., U-net~\citep{ronneberger_2015_MICCAI},  $\classifier_\theta$: $\imagespace \rightarrow \labelspace$, where $\theta$ denotes the network parameters, such as weights and biases in the convolutional layers. Assume we have a \emph{small} labeled dataset $\dataset_L: \{(\image_i
, \segmentation_i)\}_{i=1}^{N}$ ($N>0$) and an unlabeled dataset $\dataset_U: \{(\image_i)\}_{i=1}^{M} (M\geq 0)$, where images $\image$ and pixel-wise labels $\segmentation$ are drawn from the joint distribution $P(\imagespace,\labelspace)$. The learning goal is to train a network $\classifier_{\theta}$ parameterized by $\theta$ to model the conditional probability distribution $P(\labelspace|\imagespace)$. As the network usually contains millions of parameters, it is important to regularize the network to alleviate the over-fitting problem, especially when there is \emph{insufficient} training data.

Recent works on image classification have shown that consistency regularization with data augmentation can be an effective approach to regularize the network and exploit the value of unlabeled data~\citep{Xie_2020_NIPS}. Our method follows this learning paradigm. At a high level, the learning objective for the network can
be formulated as following:

\begin{equation}
\label{eq:general loss}
\min_{\theta} \EX_{\image \in \dataset_l} \Loss_{s}\left(\segmentation, \classifier_\theta(\image) \right)+\lambda \EX_{\image \in \dataset_l \cup \dataset_u} \RegLoss(\image; \classifier_\theta,\augment_\transformation).
\end{equation}

Here $\Loss_{s}$ denotes the supervised loss (e.g., cross-entropy loss) for labeled images in the training set; $\RegLoss$ is a consistency regularization term computed on both labeled and unlabeled data; $\lambda$ is a weighting factor to balance the supervised and regularization 
loss terms. In particular, $\RegLoss$ measures the inconsistency between the prediction for the original image $\classifier_\theta(\image)$ and the prediction for the image under a single or a composite perturbation/transformation function: $\classifier_\theta(\augment_\transformation(\image))$. $\augment_\transformation $ is short for $\augment(\cdot,\transformation)$ denoting the transformation function $\augment$ parameterized by $\transformation$. 

% A typical implementation with this learning framework is to apply random Gaussian noise for consistency regularization~\citep{cui_2019_semi_noise,tarvainen_2017_mean}. In this case, $\augment_\transformation$ is an image intensity perturbation function with additive Gaussian noise and one can use a distance loss function, e.g., mean squared error (MSE) loss to measure the inconsistency between two predictions (predictions before and after applying noise). In this work, we propose to use a composite distance loss function to better quantify the prediction inconsistency, which will be introduced in Sec.~\ref{SEC:consistency loss}.

\subsection{\advchain: A generic adversarial data augmentation framework for effective consistency regularization}
\label{sec:3.1}
% Different from most existing perturbation-based learning algorithms with the above framework only consider to repeatedly sample transformation parameters $\transformation$ in a random fashion~\citep{cui_2019_semi_noise,tarvainen_2017_mean}, 
In this work we employ an adversarial training approach to first optimize transformation parameters $\transformation$ so that augmented images can better regularize network training~\citep{Miyato_2018_PAMI_VAT}. In other words, we would like to first find perturbations/transformations to which the current segmentation model is most sensitive and then utilize them for consistency regularization. \revise{Different from existing adversarial data augmentation methods~\citep{chen_2020_realistic,Miyato_2018_PAMI_VAT} where they only consider a \emph{fixed} type of pixel-level perturbations, we propose \advchain, a generic adversarial data augmentation framework which allows to optimize the parameters in a random chain of different image transformation functions (incl. photo-metric and geometric transformations) for improved data diversity, with the aim of better reflecting complex image variations in MR imaging. In this work, we employ four different image transformation functions as a proof of concept (will be introduced in Sec.\ref{SEC:ADV Tr}), allowing to generate not only pixel-level perturbations but also geometric variations, e.g. morphological variations.}

The whole learning procedure can be generally described as a two-step optimization, as shown in Fig.~\ref{fig:framework}: 
\begin{itemize}
    \item With the segmentation parameters $\theta$ fixed, we update the image transformation parameters $\transformation$ in the search space to produce an adversarial image augmentation  $\augment$, so that it \textbf{maximizes} the disagreement (measured by $\RegLoss$) between the original prediction $\prediction=\classifier_\theta(\image)$ and the  prediction on the perturbed image  $\prediction'=\classifier_\theta(\augment_\transformation(\image))$. Here $\augment$ can be a single transformation or a composite transformation with chained image transformation functions $\augment: \augment_1 \circ \augment_2,...,  \circ \augment_n$ ($n>=1$) with corresponding transformation parameters $\transformation: \transformation_1, \transformation_2,..., \transformation_n$;
     \item With the optimized transformation parameters $\transformation^*$ fixed, we obtain an \advchain\ augmented image $\image^{*} = \augment_{\transformation^{*}}(\image)$ and feed it to the network to update the segmentation network parameters $\theta$ to \textbf{minimize} the supervised loss $\Loss_s$ and the consistency regularization loss $\RegLoss$. 
\end{itemize}

Mathematically, the learning objective can then be formulated as follows:
\revise{
\begin{subequations}
\label{opt:general adversarial training}
\begin{align}
&\min_{\theta} \EX_{\image \in \dataset_l} \Loss_{s}\left(\segmentation, \classifier_\theta(\image) \right)+\lambda \EX_{\image \in \dataset_l \cup \dataset_u} \RegLoss(\image; \classifier_\theta,\augment_{\transformation^*}).\\
& s.t.\; \transformation^* = 
\;\; \underset{\transformation: \condition(\transformation)}{\argmax}
\;\;\RegLoss(\image;\classifier_\theta,\augment_\transformation)
.
\end{align}
\end{subequations}}
Here, $\condition(\transformation)$ denotes a set of constraints that specify the search space of corresponding transformation parameters. These constraints are essential as they explicitly ensure that augmented or perturbed images remain meaningful and realistic. \revise{Since it is difficult to determine the optimum parameters $\transformation^*$ in practice, we relax the objective in Eq.~\ref{opt:general adversarial training}(b). We instead try to find a relatively effective $\transformation^*$ that produces higher inconsistency loss $\RegLoss$ to strengthen the network regularization, compared to its random initialized counterpart. To achieve the goal, we employ the commonly used projected gradient descent (PGD) algorithm~\citep{Madry_2017_PGDattack} to update the randomly initialized transformation parameters in a chain, which has been found effective to optimize the parameters with constraints across many applications~\citep{Xing_2021_NIPS_2021}:
\begin{equation}
\label{eq:PGD}
    \transformation_{i} \leftarrow \proj_{\condition} \; \transformation_{i} +\alpha_{i} \nabla_{\transformation_{i}} \RegLoss/\|\nabla_{\transformation_{i}} \RegLoss\|_2.
\end{equation}
Here, $\proj$ is the projection operation that projects the updated parameters onto the feasible set constrained by $\condition$, $\alpha_i$ specifies the step size when we update the parameters $\transformation_i$ for the transformation function $\augment_i$ in a chain along the direction of the normalized gradient $ \nabla_{\transformation_{i}} \RegLoss/\|\nabla_{\transformation_{i}} \RegLoss\|_2$. We apply the chain rule to efficiently compute the gradients along the augmentation chain~\footnote{Applying the chain rule allows us to calculate the gradient of the loss function with respect to the parameters of each transformation function in a chain in an efficient way. The transformation functions are required to be differentiable. In our work, all transformations satisfy the criterion, where the geometric transformations are implemented using the differentiable spatial transformer module~\citep{Max_Jaderberg_2015_NIPS_Spatial_Transformer}.}. We use normalized gradients to update the parameters in each transformation function to avoid gradient explosion or vanishing problem when the length of chained transformations is long. } 

\subsection{Increasing the data diversity of \advchain\ with dynamic transformations}
\label{sec:Enhancing consistency regularization by chaining transformations}
\advchain\ allows to optimize \emph{dynamic} transformations: e.g. single or composite transformations randomly generated at training. Such flexibility is highly adorable as the data diversity can be largely increased at a low cost. While it is possible that better performance can be achieved by employing the optimum combinations of transformation functions for a particular task, it often requires extraordinary high computational costs to search for improved data augmentation policies~\citep{Cubuk_2018_AutoAugment}. Therefore, in \advchain\, we simply randomly select and chain the transformations in an arbitrary order, allowing itself to explore all possible solutions as a trade-off between efficiency and effectiveness.  Specifically, for each image, each transformation function is randomly selected with a probability of $p$ and then chained in a random order to produce a high diversity of augmented images. We then apply adversarial training to this chain, which optimizes the underlying parameters in each transformation. 

In Algorithm~\ref{alg1}, we illustrate the detailed steps of the proposed adversarial data augmentation method with a random chain of transformations for consistency regularization. For ease of understanding, we use subscripts $1$, $2$, $3$ to represent three different arbitrary image transformation functions.
\begin{algorithm}[H]
\caption{\advchain\ }
\label{alg1}
\begin{algorithmic}[1]
 \STATE {\bfseries Input:}  labelled or unlabelled training set: $\dataset_L \cup \dataset_U$, a segmentation network  $\classifier_\theta$
    \STATE {\bfseries Requires:} a set of predefined transformations ${\augment}$, number of update steps $k$, step size $\alpha$.
    \FOR {\revise{$\image \in \dataset_L \cup \dataset_U$}}
    \STATE Randomly select transformation operations with a probability of $p$ from a group of transformation functions $\{{\augment}\}$, e.g.,  $\augment_1,\augment_2,\augment_3$.
    \STATE Chain them in a random order, e.g., $\augment_{312}=\augment_{3} \circ \augment_{1} \circ \augment_{2}$ and randomly initialize the transformation parameters  $\transformation^{(0)}_{312}:\transformation^{(0)}_3, \transformation^{(0)}_1,\transformation^{(0)}_2$
    \FOR {j=0,..., k-1}
    \STATE Compute consistency loss $\RegLoss_{
    }(\image;\classifier_\theta,\augment_{312}(\cdot;\transformation^{(j)}_{312}))$
   \STATE Apply the chain rule to computing gradients  $\nabla_{\transformation_3}\RegLoss_{
}$, $\nabla_{\transformation_1}\RegLoss_{
}$, $\nabla_{\transformation_2}\RegLoss_{
}$ and update $\transformation_3$, $\transformation_1$, $\transformation_2$, respectively using Eq.\ref{eq:PGD}
    \ENDFOR
\STATE Return the chain of data augmentation with optimized parameters to augment images: $\augment_{312}^{adv}(\image)= \augment_{312}(\image;\transformation^{adv}_{312})$.
\STATE Compute the loss for network optimization using Eq.~\ref{opt:general adversarial training}(a).
\ENDFOR

\end{algorithmic}
\end{algorithm}

\subsection{Realistic image transformation functions}
\begin{figure*}[!ht]
    \centering
    \includegraphics[width=\textwidth]{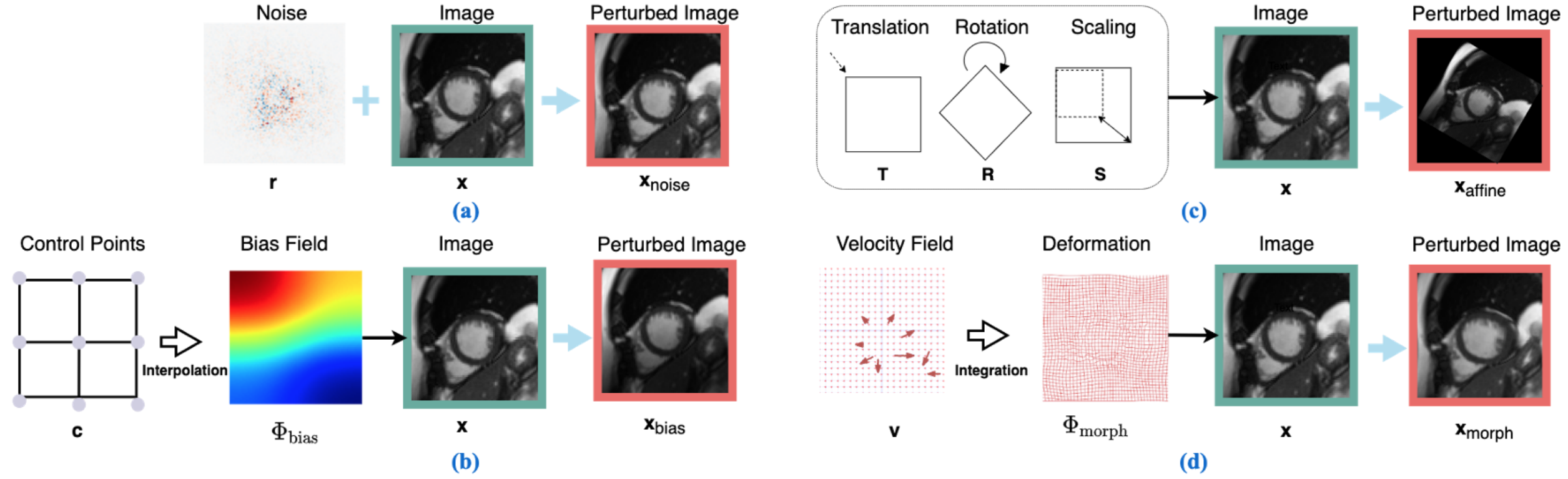}
    \caption{Adversarial example construction with: (a) image noise model $\augment_{\rm noise}$: (b) image intensity transformation model with bias field $\augment_{\rm bias}$; (c) image geometric transformation model $\augment_{\rm affine}$; (d) image deformation model $\augment_{\rm morph}$. Best viewed in color.}
    \label{fig:transformation_family}
\end{figure*}
\label{SEC:ADV Tr}
\revise{\advchain\ is  an advanced, generic, lightweight adversarial data augmentation framework, which can be applied to optimize any types of differentiable photometric and geometric transformations on-the-fly.}
In this work, we consider four different image transformation functions as a proof of concept. The transformation functions are constructed to reflect common data variations that exist in MR images, including:
\begin{itemize}
    \item an additive image noise model $\augment_{ \rm noise}$, which adds pixel-wise noise to images (Sec.~\ref{SEC:ADV noise}),
    \item an image intensity transformation model $\augment_{\rm {bias}}$, which generates bias fields to corrupt images. Bias field can introduce intensity inhomogeneities in images, which is a common artefact in MR imaging~\citep{sled_1998_TMI_nonparametricN3ITK,Tustison_2010_N4ITK,Ferreira_2013_JCMR_Artefacts}(Sec.~\ref{SEC:bias field}),
     \item an image geometric transformation model $\augment_{\rm{affine}}$, which simulates image spatial variance caused by patient movement and the adjustment of acquisition parameters (e.g., image resolution, field-of-view) during scanning (Sec.~\ref{SEC:affine}),
   \item a {diffeomorphic deformation model} $\augment_{\rm{morph}}$, which simulates inter- and intra-subject shape variability attributed to age, pathology, and motion (Sec.~\ref{SEC:morph}).
\end{itemize}

Without loss of generality, we assume that all image transformations are defined in 2D space and performed on 2D images $\image \in \Reals^{H\times W}$. One should note that these transformation can be potentially extended to 3D space.

\subsubsection{Image noise model \texorpdfstring{$\augment_{\rm noise}$}{Tnoise}}
\label{SEC:ADV noise}
We begin with the most commonly used image noise model, which applies additive noise to perturb images, as shown in Fig.~\ref{fig:transformation_family}(a). Following existing adversarial data augmentation works ~\citep{Goodfellow_2015_FGSM,Madry_2017_PGDattack,Miyato_2018_PAMI_VAT}, the image noise model is formulated as
\begin{equation}
\image_{\rm{noise}}=\augment_{\rm noise}(\image;\noise)=\image+\noise,
\end{equation}
where the magnitude of the noise $\noise$ is bounded by $\epsilon_{noise}$. The magnitude constraint $\condition_{noise}$ can be described as follows:
\begin{equation}
\label{eq:noise constraints}
 \|\noise\|_2 \leq \epsilon_{\rm noise}.
\end{equation}
Here $\epsilon_{\rm noise} (\epsilon_{\rm noise} \geq 0)$ is a scalar controlling the level of noise.

\subsubsection{Image intensity transformation with bias field \texorpdfstring{$\augment_{\rm {bias}}$}{Tbias}}
\label{SEC:bias field}
Following previous related works on bias field correction~\citep{sled_1998_TMI_nonparametricN3ITK,Tustison_2010_N4ITK}, a multiplicative intensity transformation is adopted here to introduce intensity non-uniformity to images. As shown in Fig.~\ref{fig:transformation_family}(b), the intensity of the image $\image$ is perturbed by multiplying with the bias field $\Phi_{\rm{bias}} \in \Reals^{H \times W}$:
\begin{equation}
\image_{\rm{bias}}=\biasfunc=\image \odot \Phi_{\rm bias}(\cpoints),
\end{equation}
where $\odot$ denotes point-wise multiplication. Similar to the bias field model in N4ITK~\citep{Tustison_2010_N4ITK}, we use a set of uniformly distributed $b \times b$ control points $\cpoints=\{c^{(i,j)}\}_{b \times b}$ for bias field construction, based on the fact that the bias field is smoothly varying across the image, see Fig.~\ref{fig:transformation_family}(b). Following~\citet{sled_1998_TMI_nonparametricN3ITK,Tustison_2010_N4ITK}, $\cpoints$ is defined in the log-transformed bias field space for numerical stability during optimization.  A smooth bias field is obtained by first interpolating a grid of regularly spaced control points $\cpoints$ with a third-order B-spline smoothing~\citep{Gallier_2000_curves_book} and then taking the exponential for value recovering: $\Phi_{\rm {bias}}(\cpoints)=\exp( \textit{B}(\cpoints))$. Here, \textit{B} represents the interpolation function with uniform B-splines for smoothing.
When $\cpoints=\0$, $\Phi_{\rm {bias}}=\one$ (identity field).

The magnitude constraint for the bias field perturbation $\condition_{\rm bias}$  is defined as:
\begin{equation}
\label{eq:bias field constraints}
\forall{(x,y) \in \Reals^2},
\|{\Phi_{\rm bias}}(\cpoints) -\one\|_\infty \leq \epsilon_{\rm {bias}},
\end{equation}
where $\epsilon_{\rm {bias}} (\epsilon_{\rm {bias}} \geq 0)$ is a scalar to control the maximum level of corruption caused by the bias field.

\subsubsection{Image geometric transformation model \texorpdfstring{$\augment_{\rm{affine}}$}{Taffine}}
\label{SEC:affine}
To model image-level geometric variations we use an affine transformation to transform images. This process is defined as:
\begin{equation}
\image_{\rm affine}=\augment_{\rm{affine}}(\image;\mathbf{a})=\augment_{\rm{affine}}(\image;t_x, t_y, r, s_x,s_y),
\end{equation}
where $\mathbf{a}$ contains five affine parameters $[t_x, t_y, r, s_x,s_y]$ to characterize  translation $T$, rotation $R$ and scaling $S$ operations which are performed in succession, see Fig.~\ref{fig:transformation_family}(c).  Given a 2D image $\image$ and the affine parameters $\mathbf{a}$, a pixel $\image{(u,v)}$ at position $(\xpos, \ypos)$ in the original image $\image$ is transformed to a new position $(\xpos',\ypos')$ via the following matrix multiplications:
\begin{equation}
\left[\begin{array}{l}\xpos'\\ \ypos' \\ 1\end{array}\right] =(T\cdot R\cdot S)  \cdot\left[\begin{array}{l}\xpos\\ \ypos \\ 1\end{array}\right], 
\end{equation}
where 
$T=\left[\begin{array}{ccc}1 & 0 & t_{x} \\ 0 & 1 & t_{y} \\ 0 & 0 & 1\end{array}\right]$,
$R = \left[\begin{array}{ccc}\cos r\pi & -\sin r\pi & 0 \\ \sin r\pi & \cos r\pi & 0 \\ 0 & 0 & 1\end{array}\right]$, $S=\left[\begin{array}{ccc} 1+s_{x} & 0 & 0 \\ 0 & 1+s_{y} & 0 \\ 0 & 0 & 1\end{array}\right]
$. 
We use a normalized Cartesian coordinate system centred at (0,0) to specify each pixel's location $(\xpos, \ypos)$. Each location is normalized by the input spatial dimensions so that its value lies in $[-1,1]$: $-1\leq \xpos \leq 1, -1\leq \ypos \leq 1$. Each transformation parameter is restricted in a user-defined range to control the range of the spatial transformations. The constraint for the affine transformation model $\condition_{affine}$ can be described as:
\begin{equation}
\label{eq:affine constraints}
-\epsilon_{a^{i}} \leq a^{i} \leq \epsilon_{a^{i}}; \forall a^i \in [t_x, t_y, r, s_x,s_y].
\end{equation} 

% When $\mathbf{a} =\0$, this transformation is equal to the identity mapping $\identity$. 

% \textbf{Inverse affine transformation $\augment_{affine}^{-1}$:}
% Since the affine transformation is parameterised by a homogeneous transformation matrix, its inverse transformation matrix can be directly computed via: $\augment_{\rm affine}^{-1} = S^{-1}R^{-1}T^{-1}$ .

\subsubsection{Image deformation model \texorpdfstring{$\augment_{\rm morph}$}{Tmorph}}
\label{SEC:morph}
% \todo{reduce the length of morphological transformation}
{To introduce intra- and inter-subject anatomical variations, we would like to construct a generator which can produce a smooth spatial transformation to deform the image, while preserving its topology and spatial layout. To achieve the goal, we model these variations using invertible and differentiable diffeomorphic transformations~\citep{Vercauteren_2009_NeuroImage_Demons}. Specifically, following previous works on diffeomorphic image registration, e.g., Demons~\citep{Vercauteren_2009_NeuroImage_Demons}, we parameterize the deformation $\Phi_{\rm{morph}}$ using an underlying stationary velocity field $\velocity$ that
$
\frac{\partial{{\Phi_{\rm{morph}}}(t)}}{\partial{t}}=\velocity ( \Phi_{morph}^{(t)}), 
$
where $\Phi_{morph}^{(t)}$ represents the deformation at time $t$. The final deformation ${\Phi_{\rm{morph}}}$ is obtained by starting with an identity transform $\Phi_{morph}^{(0)}=\textit{Id}$ and integrating the stationary velocity field $\velocity$ over $t \in [0,1]$\footnote{We employ the \emph{scaling and squaring}~\citep{arsigny_2006_MICCAI_log} to approximate the integration to accelerate the computation as a common practice~\citep{Balakrishnan_2019_TMI_VoxelMorph,arsigny_2006_MICCAI_log,Vercauteren_2009_NeuroImage_Demons}.}.
As shown in Fig.~\ref{fig:transformation_family}(d), given a 2D image $\image$ and a 2-dimensional velocity field $\velocity$, the deformed image is obtained using the following function:
\begin{equation}
\image_{\rm{morph}}=\difffunc=\image \circ {\Phi_{\rm{morph}}} = \image \circ \int_{t=0}^{1}\velocity( \Phi_{\rm {morph}}^{(t)}) \; dt.
\end{equation}
Here $\image \circ {\Phi_{\rm{morph}}}$ represents $\image$ warped by a deformation field ${\Phi_{\rm{morph}}} \in \Reals^{H\times W \times 2}$. 
To initialize the velocity field $\velocity$, we sample a random, low-resolution tensor $\velocity' \in \Reals^{\frac{H}{ds} \times \frac{W}{ds}\times 2}$ $(ds\geq 1)$ \footnote{We apply bilinear upsampling to $\velocity'$ to obtain $\velocity$.}. We impose a magnitude constraint $\condition_{\rm{morph}}$ to  $\velocity'$ to control the level of deformation:
\begin{equation}
\label{eq:velocity constraints}
 \|\velocity' \|_2 \leq \epsilon_{\rm{morph}}.
\end{equation}}
This is achieved by directly applying $\rm{L}2$ norm to $\velocity'$ and re-scaling it to find an approximate solution in the constrained space: $\velocity' \leftarrow \epsilon_{\rm{morph}} \frac{\velocity'}{\|\velocity'\|_2}$~\citep{Miyato_2018_PAMI_VAT}. 
To further encourage the spatial smoothness of the deformation, following the related work on the diffeomorphic demons \citep{Vercauteren_2009_NeuroImage_Demons}, we apply Gaussian smoothing $K_{\rm smooth}$ to the velocity field: $\velocity' \leftarrow K_{\rm smooth} (\velocity')$ as well as to the integrated deformation: $\Phi_{\rm smooth} \leftarrow K_{\rm smooth} (\Phi_{\rm morph})$. In this way, we ensure the deformation is smooth and diffeomorphic without introducing additional smoothness regularization terms, simplifying the optimization procedure~\citep{cachier_2003_iconic}. In the experiments, we used a small Gaussian kernel $ K_{\rm smooth}$ with $\sigma_{smooth}=1$, as suggested by~\citet{Vercauteren_2009_NeuroImage_Demons}.

\subsection{Consistency loss function \texorpdfstring{$\RegLoss$}{R}} 
\label{SEC:consistency loss}
\subsubsection{Consistency loss function for photometric transformations} 
{For photometric transformations, i.e. $ \augment_{\rm{bias}}$, $\augment_{\rm noise}$, we directly use a composite distance loss function $\Distance$ to compute the consistency regularization term  $\RegLoss$ computed on the original probabilistic prediction $\classifier_\theta(\image)$ and perturbed prediction $\classifier_\theta(\augment(\image))$:
\begin{equation}
\RegLoss_{\Distance}(\image;\classifier_\theta,\augment) =\Distance(\classifier_\theta(\image) \; ,\classifier_\theta(\augment(\image;\transformation))).
\label{eq:invariance loss}
\end{equation}
The composite distance function $\Distance$ measures two predictions $\p,\p'$ in the same image coordinates, which is defined as follows:
\begin{align}
\begin{split}
\Distance(\p,\p') & = \Loss_{MSE}(\p,\p')+w \Loss_{Contour}(\p,\p'), \\
\Loss_{MSE}(\p,\p') & = \|\p-\p'\|_2^2, \\
\Loss_{Contour} &=  \sum_{c \neq BG}^{C}\sum_{S \in \ {S_x,S_y}}\|S(\p^{(c)})-S(\p'^{(c)})\|_2^2.
\end{split}
\end{align}
Here we adopt the mean-squared-error loss $\Loss_{MSE}$ to measure pixel-wise differences, as a common practice in consistency regularization related works~\citep{tarvainen_2017_mean,Li_2020_TNNLS_transformation, cui_2019_semi_noise,berthelot_2019_mixmatch}. In addition, we employ a contour-based loss function $\Loss_{Contour}$ to better capture the difference on the foreground objects' boundaries between two predictions~\citep{Chen_2019_STACOM_unsupervised}. $S_x,S_y$ represent Sobel filters in the x- and y- directions, which are used to extract object boundaries from model's probabilistic map for every class $c$ except the background (BG) class. $w$ is a weight that controls the relative importance of two terms. In our experiments, we empirically set it to $0.5$. We believe that combining pixel-wise and contour-based loss terms can help the network to better capture the semantic dissimilarity between two predicted segmentation maps.
}
\subsubsection{Consistency loss function for geometric transformations}
For geometric transformations, i.e., $ \augment_{\rm{morph}}$, $\augment_{\rm{affine}}$, Eq.~\ref{eq:invariance loss} is not directly applicable as the position and/or structural information of target objects also changes accordingly. We therefore transform the perturbed prediction back to the coordinates of the original image accordingly before computing the consistency loss. The regularization loss is defined as:
\begin{equation}
\RegLoss_{\Distance}(\image;\classifier_\theta,\augment) = \Distance(\classifier_\theta(\image),\augment^{-1}_\transformation(\classifier_\theta(\augment(\image;\transformation)))).
\label{eq:equivariance loss}
\end{equation}
Here $\augment_\transformation^{-1}$ denote the inverse transformation for $\augment(\cdot;\transformation)$.
The inverse transformations for the two types of geometric transformations are easy to compute:
\begin{itemize}
    \item \textbf{Inverse affine transformation} $\augment_{\rm affine}^{-1}$:
            Since the affine transformation is parameterised by a homogeneous transformation matrix, its inverse transformation matrix can be directly computed via: $\augment_{\rm affine}^{-1} = S^{-1}R^{-1}T^{-1}$;
    \item \textbf{Inverse deformation} $\augment_{\rm morph}^{-1}$:
            The inverse deformation $\Phi_{\rm morph}^{-1}$ is obtained by integrating the \emph{negative} velocity field ($-\velocity$) backward: $\augment_{\rm morph}^{-1} =  \int_{t=0}^{-1}(-\velocity) (\Phi^{(t)}) dt$~\citep{Ashburner_2007_diffeomorphic}. 
\end{itemize}
\subsubsection{Consistency loss function for a chained transformation} 
{
For a chained transformation $\augment_{1 \circ 2 \circ ... \circ m}:\augment_{1} \circ \augment_{2} \circ ... \augment_{m}$ including both photometric and geometric transformations, we employ Eq.~\ref{eq:invariance loss} and Eq.~\ref{eq:equivariance loss} to compute the consistency loss between the original prediction and the perturbed prediction:
\begin{equation}
\label{eq:adv chain loss}
\RegLoss_{\Distance}^{chain}= \RegLoss_{\Distance}(\image;\classifier_\theta,\augment_{1 \circ 2 \circ ... \circ m}).
\end{equation}
This means one needs to transform the perturbed prediction back to the coordinates of the original image if there is any geometric transformation involved. For instance, given a chain of transformation functions $\augment_{\rm affine \; \circ \; \rm noise}$ : $\augment_{\rm affine} \circ \augment_{\rm noise}$, the loss function is defined as follows:
\begin{equation}
\RegLoss^{chain}_{\Distance}(\image;\classifier_\theta,\augment_{\rm affine \; \circ \; \rm noise})
= \Distance(\prediction,\augment_{\rm affine}^{-1}(\prediction'))
\end{equation}
where $\prediction=\classifier_\theta(\image), \prediction'=\classifier_\theta(\augment_{\rm affine \; \circ \; \rm noise}(\image))$. }

\section{Experiments Settings}
\label{SEC: experiment}
\subsection{Datasets}
\label{SEC:dataset}
\subsubsection{Cardiac MR dataset} The cardiac dataset is provided by The Automated Cardiac Diagnosis Challenge (ACDC)~\citep{Bernard_2018_ACDC}~\footnote{\url{https://www.creatis.insa-lyon.fr/Challenge/acdc/databases.html}}, which is a public dataset for cardiac MR image segmentation. The left ventricular cavity (LV), the left ventricular myocardium (MYO), and the right ventricular cavity (RV) in end-diastolic and end-systolic frames were manually labeled by experts. The original in-plane pixel spacing ranges from $1.37 \times 1.37\; mm^2$ to $1.68 \times 1.68\; mm^2$.  %This dataset was collected from 100 subjects, which were evenly classified into 5 populations: 1 normal group (NOR) and 4 pathological groups with cardiac abnormalities: dilated cardiomyopathy (DCM); hypertrophic cardiomyopathy (HCM); myocardial infarction with altered left ventricular ejection
%fraction (MINF); abnormal right ventricle (ARV). 

We preprocessed images to have the same in-plane pixel spacing:$1.37 \times 1.37\; mm^2$, following~\citet{Chaitanya_2019_IPMI}. After that, all images were centrally cropped to $192 \times 192$ in order to save computational cost. We used the same data setting as in \citet{Chaitanya_2019_IPMI}, splitting the dataset (100 subjects in total) into 4 subsets: an unlabeled set for semi-supervised learning ($M$=25), a validation set (5 subjects) and a test set (20 subjects). \revise{The rest 50 subjects were used as the labeled training pool.} We selected $N$ subjects from the rest to form a  labeled set for training, simulating a low-data learning regime. Specifically, we evaluated  one-shot learning ($N$=1) and three-shot learning ($N$=3) in both supervised (using the labeled set only) and semi-supervised (using both labeled and unlabeled sets) settings. \revise{We also trained the segmentation with different numbers of labeled subjects from the pool (N=10, N=25) to test the performance improvements against different settings. In all settings, we trained the network for five times, each time with a different, randomly selected labeled set to alleviate the dataset selection bias, and reported the mean performance.}

\subsubsection{Prostate MR dataset}
The prostate dataset is provided by the Medical Segmentation Decathlon Challenge~\citep{Antonelli2021_Decathlon}\footnote{\url{http://medicaldecathlon.com/}}, which consists of 32 subjects. The peripheral zone (PZ) and the central zone (CZ) of the prostate have been manually labeled and verified by an expert human rater. We performed segmentation on T2 images, where all images have been resampled to have the same pixel spacing ($0.625 \times 0.625 \;mm^2$, the median value of pixel spacings in this dataset) and then centrally cropped to $224 \times 224$ to reduce computational cost.

To train and evaluate the proposed method, we split the dataset into 22/4/6 for training/validation/testing. The training set was further divided into two subsets (11 subjects each). We randomly selected $N \; (\mbox{with}\; N\leq11)$ subjects from the first set to form a small labeled set while all subjects in the second one were used to construct the unlabeled set $(M=11)$ for semi-supervised learning. We trained the network for three times, each time with a different, randomly selected labeled set, and reported the mean performance.

\subsection{Implementation details}
\label{SEC: implementation details}
\subsubsection{Default data augmentation}
For all experiments, we applied a random data augmentation pipeline as a default setting. This augmentation pipeline includes random affine transformation (i.e. scaling, rotation, translation), image flipping, random global intensity transformation (brightness and contrast), and elastic transformation. Detailed configurations of these random transformations can be found in~\citet{Chaitanya_2019_IPMI}. After random data augmentation, the image intensity was rescaled to $[0,1]$.
\subsubsection{Training details}
The proposed method is independent of the network structures. \revise{For ease of comparison, we adopted the commonly-used 2D U-net~\citep{ronneberger_2015_MICCAI} as our segmentation network, which has been demonstrated its superiority across various medical image segmentation datasets~\citep{Isensee2021-cn}.}
The Adam optimizer was used to update network parameters with a batch size of 20. To accelerate training, we first trained the network with the default data augmentation for 1,000 epochs (learning rate=$10^{-3}$) to get a pre-trained segmentation network. We then fine-tuned the network with the proposed data augmentation method. We applied a small learning rate ($10^{-5}$) and trained the network for 600 epochs. Following previous work on semi-supervised learning~\citep{berthelot_2019_mixmatch,tarvainen_2017_mean,Li_2020_TNNLS_transformation}, we evaluated the segmentation model using an exponential moving average of model parameters with a decay rate of 0.999.\\
\textbf{Loss configuration}:
For the supervised loss $\Loss_{s}$, we used a combination of a weighted cross entropy loss function and a soft Dice loss function~\citep{Baumgartner_2018_STACOM}, to alleviate the class imbalance problem in data. Empirically, for cardiac segmentation, class weights for background (BG), LV, MYO, RV were set to 0.01:0.33:0.33:0.33 respectively to give equal weights to foreground classes; for prostate segmentation, weights for BG, PZ, CZ were set to 0.01:0.66:0.33. We set a higher weight to the PZ class, as it has significantly fewer number of pixels in each image, compared to the CZ class.  Regarding the consistency regularization  loss $\RegLoss$, 
the weighting parameters for the contour loss term $w$ was empirically set to 0.5. \revise{Since the prediction for images $\classifier_\theta({\image})$ can be very noisy at the beginning of the training, it can produce incorrect supervision signal to misguide the training. This is a common issue in consistency-regularized methods~\citet{Li_2020_TNNLS_transformation,berthelot_2019_mixmatch}. Similar to~\citet{Li_2020_TNNLS_transformation,berthelot_2019_mixmatch}, we started the training with a small weight $\lambda$ for $\RegLoss$ and slowly increased it in the first $e_{ramp}$ epochs until it reached to its maximum $\lambda^{max}$. The value of $\lambda$ was linearly increased with the number of training epochs: $\lambda=\min(\lambda^{max}\times e/e_{ramp},\lambda^{max})$. $e$ is the number of the current epoch and $e_{ramp}=200$. We empirically set $\lambda^{max}=1.0$ so that in the later stage the supervised loss $\Loss_{s}$ and the consistency regularization loss $\RegLoss$ share the same weight to balance the training. Ideally, the two losses are expected to be zero when the network reaches to its optimum, suggesting that the network not only produces correct segmentations on original images but also produces consistent predictions on adversarially augmented images.}

\textbf{Adversarial data augmentation configuration:} For bias field construction, we adopted the B-spline convolution kernel provided by the AirLab library to interpolate the control points~\citep{Sandkuhler_2018_Arxiv_Airlab}. \revise{To ensure the generated variations to be realistic, one has to specify the magnitude constraints for each transformation. For simplicity, one can reuse the magnitude constraints specified in existing hand-crafted random data augmentation works or empirically set up the range based on visual inspection, which is a common
practice in most data augmentation frameworks. Specifically, in this work, the constraints for noise, bias field, rotation and translation were directly taken from our previous work~\citep{chen_2020_realistic} and RandAugment~\citep{Cubuk_2020_RandAugment} whereas for our proposed novel image deformation model, the velocity magnitude constraint was set based on visual inspection. We provide an interactive Jupyter notebook for readers' interest to visualize the augmented images with different transformation configurations in our code repository~\footnote{\url{https://github.com/cherise215/advchain/tree/master/example}}. Table~\ref{tab:transformation_parameter} lists the detailed configurations for the employed transformations in our work. The probability of selecting each transformation $p$ is set to $0.5$ for simplicity.} 
\begin{table}[!ht]
\centering
\caption{Configurations for image transformations}
\label{tab:transformation_parameter}
\resizebox{\columnwidth}{!}{%
\begin{tabular}{@{}lll@{}}
\toprule
Transformation & \multicolumn{2}{l}{Parameter constraints} \\ \midrule
$\augment_{\rm noise}$ & \multicolumn{2}{l}{noise $\|\rm \noise\|_2 \leq \epsilon_{\rm noise}=1$} \\\midrule
$\augment_{\rm bias}$ & \multicolumn{2}{l}{control points $\cpoints \in \Reals^{b \times b}$; $b=4, \epsilon_{\rm bias} =0.3$} \\ \midrule
$\augment_{\rm affine}$ & \multicolumn{2}{l}{\begin{tabular}[c]{@{}l@{}}translation: $-0.1 \leq t_x,t_y \leq 0.1$\\ rotation: $-\frac{30^\circ}{180^\circ} \leq r \leq \frac{30^\circ}{180^\circ}$\\ scaling:$-0.2 \leq s_x,s_y \leq 0.2$\end{tabular}} \\\midrule
$\augment_{\rm morph}$ & \multicolumn{2}{l}{$\velocity' \in \Reals^{\frac{H}{ds} \times \frac{W}{ds} \times 2}$, $ds=16$, $\|\velocity'\|_2 \leq \epsilon_{\rm morph}=1.5$}  \\ \bottomrule
\end{tabular}%
}
\end{table}

\revise{For the optimization of the underlying transformation parameters, we first randomly sampled the parameters from the specified range for initialization and then applied projected gradient descent to ensure the updated transformations are still within the search space. Specifically, for noise $\noise$ and velocity fields $\velocity$ with $l^2$ {norm} constraints, similar to \citep{Miyato_2018_PAMI_VAT}, we normalized and re-scaled the updated parameters to meet the magnitude constraints specified in Eq.~\ref{eq:noise constraints} and Eq.~\ref{eq:velocity constraints}, respectively. For the bias field, we clipped the values of generated bias field to meet the criterion specified in Eq.~\ref{eq:bias field constraints}. For affine transformation, we applied the element-wise HardTanh activation function to the transformation parameters (rotation, translation, scaling) and re-scaled them to meet the criterion specified in Eq.~\ref{eq:affine constraints}. We used the same step size ($\alpha_i=1$) and performed only one-step ($k=1$) search for simplicity and training efficiency, similar to ~\citet{Miyato_2018_PAMI_VAT}. Detailed implementation can be found in our code repository~\footnote{\url{https://github.com/cherise215/advchain/tree/master/advchain/augmentor}}. We used the same configuration for both cardiac and prostate segmentation tasks to test the generality. Results show that it can yield substantial improvements for both applications. 
}
The full code implementation for \advchain\ is based on PyTorch and is available at GitHub~\footnote{\url{https://github.com/cherise215/advchain}}. All experiments were performed on an Nvidia$^{\tiny{\text{\textregistered}}}$
GeForce$^{\tiny{\text{\textregistered}}}$ 2080 Ti. 

% We did not perform a grid search to find an optimal one for each application, as it will be too time-consuming to be practical.The parameters used in this work are supposed to generalize across different applications. Yet, researchers are welcomed to tune these transformation parameters for improved performance in their applications. We have to point out the main aim of this paper is not to find the best data augmentation policy but to find effective transformation parameters (still within a given policy) for a particular image during the network training.

\section{Results}
\subsection{Comparison study}
\begin{table*}[t]
\centering
\caption{\revise{
Comparison between the proposed method (\advchain) against high-performing consistency regularized semi-supervised learning methods on the cardiac, and prostate segmentation datasets. Reported values are mean Dice scores for each class. We also report the mean and standard deviation of average Dice scores over all foreground classes (AVG) for each task. N: $\#$ of labeled images, M:$\#$ of unlabeled images. LV: left ventricle; MYO: left ventricular myocardium; RV: right ventricle; PZ: peripheral zone; CZ: central zone.}}
\label{tab:SSL_comparison}
\resizebox{\textwidth}{!}{%
\begin{threeparttable}
\begin{tabular}{@{}lcccccccc|cccccc@{}}
\toprule
Task & \multicolumn{8}{c|}{Cardiac} & \multicolumn{6}{c}{Prostate} \\ \midrule
\multicolumn{1}{l|}{Dataset setting} & \multicolumn{4}{c|}{N=1, M=25} & \multicolumn{4}{c|}{N=3, M=25} & \multicolumn{3}{c|}{N=3, M=11} & \multicolumn{3}{c}{N=11, M=11} \\ \midrule
\multicolumn{1}{l|}{Method} & LV & MYO & \multicolumn{1}{c|}{RV} & \multicolumn{1}{c|}{AVG} & LV & MYO & \multicolumn{1}{c|}{RV} & AVG & PZ & \multicolumn{1}{c|}{CZ} & \multicolumn{1}{c|}{AVG} & PZ & \multicolumn{1}{c|}{CZ} & AVG \\ \midrule
\multicolumn{1}{l|}{Pretrained} & 0.5155 & 0.4290 & \multicolumn{1}{c|}{0.2201} & \multicolumn{1}{c|}{0.3882 (0.2353)} & 0.8269 & 0.7905 & \multicolumn{1}{c|}{0.6288} & 0.7487 (0.1154) & 0.3897 & \multicolumn{1}{c|}{0.7075} & \multicolumn{1}{c|}{0.5486 (0.1034)} & 0.5077 & \multicolumn{1}{c|}{0.8019} & 0.6548 (0.0934) \\\midrule
\multicolumn{1}{l|}{MixMatch~\citep{berthelot_2019_mixmatch}} & 0.6611 & 0.5415 & \multicolumn{1}{c|}{0.3547} & \multicolumn{1}{c|}{0.5191 (0.1913)} & 0.8406 & 0.8203 & \multicolumn{1}{c|}{0.6849} & 0.7819 (0.1050) & 0.4770 & \multicolumn{1}{c|}{0.7505} & \multicolumn{1}{c|}{0.6137 (0.0701)} & 0.5893 & \multicolumn{1}{c|}{0.8220} & 0.7057 (0.0566) \\
\multicolumn{1}{l|}{FixMatch~\citep{sohn_2020_fixmatch}} & 0.6437 & 0.5496 & \multicolumn{1}{c|}{0.3666} & \multicolumn{1}{c|}{0.5200 (0.1675)} & 0.8370 & 0.8119 & \multicolumn{1}{c|}{0.6461} & 0.7650 (0.1029) & 0.4243 & \multicolumn{1}{c|}{0.7327} & \multicolumn{1}{c|}{0.5785 (0.0756)} & 0.5439 & \multicolumn{1}{c|}{0.8107} & 0.6773 (0.0726) \\
\multicolumn{1}{l|}{TCSM~\citep{Li_2020_TNNLS_transformation}} & 0.6391 & 0.5491 & \multicolumn{1}{c|}{0.3369} & \multicolumn{1}{c|}{0.5084 (0.2283)} & 0.8442 & 0.8179 & \multicolumn{1}{c|}{0.6816} & 0.7812 (0.1047) & 0.4740 & \multicolumn{1}{c|}{0.7512} & \multicolumn{1}{c|}{0.6126 (0.0753)} & 0.5986 & \multicolumn{1}{c|}{0.8248} & 0.7117 (0.0613) \\
\multicolumn{1}{l|}{VAT~\citep{Miyato_2018_PAMI_VAT}} & 0.6729 & 0.5228 & \multicolumn{1}{c|}{0.3400} & \multicolumn{1}{c|}{0.5119 (0.2218)} & 0.8598 & 0.8353 & \multicolumn{1}{c|}{0.6646} & 0.7866 (0.0999) & 0.4571 & \multicolumn{1}{c|}{0.7565} & \multicolumn{1}{c|}{0.6068 (0.0763)} & 0.5436 & \multicolumn{1}{c|}{0.8106} & 0.6771 (0.0648) \\
\multicolumn{1}{l|}{FixMatch+VAT~\citep{Wang_2021_MedIA_Deep}} & 0.6675 & 0.5778 & \multicolumn{1}{c|}{0.3715} & \multicolumn{1}{c|}{0.5389 (0.1666)} & 0.8396 & 0.8121 & \multicolumn{1}{c|}{0.6504} & 0.7674 (0.1006) & 0.4254 & \multicolumn{1}{c|}{0.7175} & \multicolumn{1}{c|}{0.5715 (0.0886)} & 0.5576 & \multicolumn{1}{c|}{0.8226} & 0.6901 (0.0642) \\ \midrule
\multicolumn{1}{l|}{\textbf{\advchain}\ (proposed)} & \textbf{0.7151} & \textbf{0.6369} & \multicolumn{1}{c|}{\textbf{0.4064}} & \multicolumn{1}{c|}{\textbf{0.5861 (0.1939)}} & \textbf{0.8708} & \textbf{0.8469} & \multicolumn{1}{c|}{\textbf{0.7072}}	&\multicolumn{1}{c|}{\textbf{0.8083 (0.0849)}} & \textbf{0.5243} & \multicolumn{1}{c|}{\textbf{0.7742}} & \multicolumn{1}{c|}{\textbf{0.6492 (0.0789)}} & \textbf{0.6245} & \multicolumn{1}{c|}{\textbf{0.8405}} & \textbf{0.7325 (0.0474)} \\ \midrule
\multicolumn{1}{l|}{Upperbound*} & 0.8963 & 0.8553 & \multicolumn{1}{c|}{0.7419} & \multicolumn{1}{c|}{0.8312 (0.0730)} & 0.8951 & 0.8627 & \multicolumn{1}{c|}{0.7654} & 0.8411 (0.0600) & 0.5930 & \multicolumn{1}{c|}{0.7960} & \multicolumn{1}{c|}{0.6945 (0.0666)} & \multicolumn{1}{l}{0.6298} & \multicolumn{1}{l|}{0.8280} & \multicolumn{1}{l}{0.7288 (0.0648)} \\ \bottomrule
\end{tabular}
\begin{tablenotes}
\item[*] {\revise{Upperbound performance of the segmentation network (U-net) when trained using labeled images (N + M) from both the labeled set and the unlabeled set. }}
\end{tablenotes}
\end{threeparttable}
}
\end{table*}
\begin{figure*}[!htbp]
    \centering
    \includegraphics[width=0.8\textwidth]{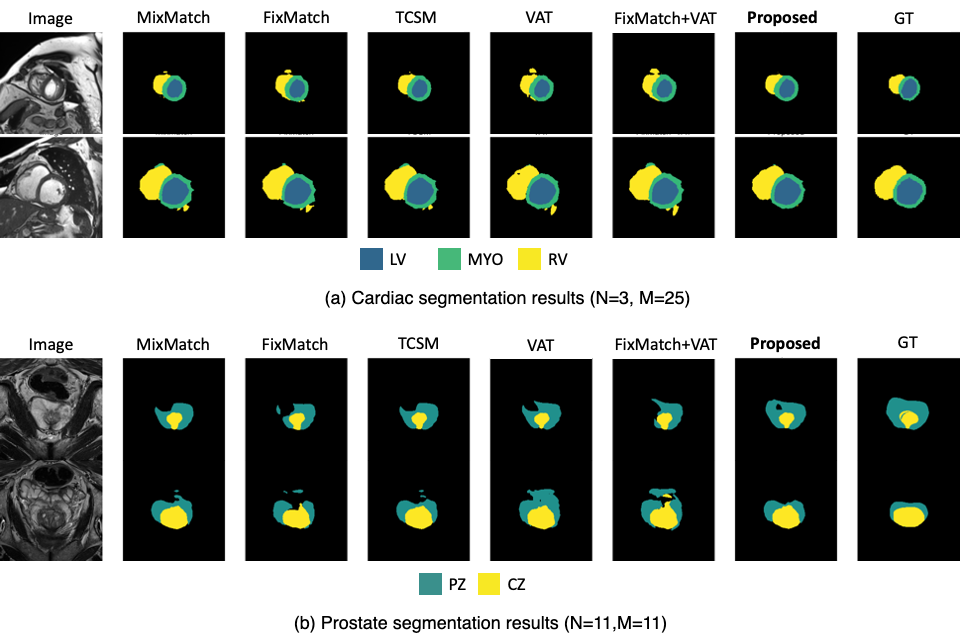}
    \caption{Visualization of the  segmentation results on the (a) cardiac and (b) prostate test data with U-net trained using different consistency regularization-based semi-supervised methods. The proposed method (\advchain) consistently outperforms the other competing methods in both tasks, producing more anatomically correct segmentation results. GT: manual labels. Best viewed in color.}
    \label{fig:SSL_comparison}
\end{figure*}

We compared our method (\advchain) to several high-performing consistency-regularization-based semi-supervised methods powered by different data augmentation techniques, which are mostly related to ours: 
\begin{itemize}
\item MixMatch~\citep{berthelot_2019_mixmatch}\footnote{\url{https://github.com/google-research/mixmatch}} is a semi-supervised learning method based on Mixup~\citep{Zhang_2018_ICLR_mixup}. Mixmatch performs linear interpolation to mix both labeled examples and unlabeled examples to get augmented image-label pairs;
\item \revise{FixMatch~\citep{sohn_2020_fixmatch}\footnote{\url{https://github.com/google-research/fixmatch}} enhances regularization by enforcing the prediction consistency between weakly augmented images (i.e., flip, shifts) and strongly augmented images with RandAugment~\citep{Cubuk_2020_RandAugment}};

%\item \footnote{To get pseudo labels on unlabeled images for data mixing, we apply label guessing. This is done by averaging predictions on $X$ randomly augmented unlabeled samples. In our experiments, $X=2$ as suggested in the original paper.}. 
% \item Mean-Teacher is a semi-supervised method which employs \emph{random noise perturbations} for consistency regularization. This method has been successfully applied to medical image segmentation.
\item TCSM~\citep{Li_2020_TNNLS_transformation}\footnote{\url{https://github.com/xmengli999/TCSM}} is an enhanced Mean-Teacher based semi-supervised learning method~\citep{cui_2019_semi_noise}. It enhances the consistency regularization by extending the noise perturbation with \emph{random geometric transformations} including scaling and rotation;
\item VAT~\citep{Miyato_2018_PAMI_VAT}\footnote{\url{https://github.com/takerum/vat_tf}} is an adversarial noise-based semi-supervised learning method. Unlike the proposed method, it injects only adversarial noise to clean data for consistency regularization and use confidence thresholding to obtain pseudo labels for reliable regularization.
% \item For Mixup, as the labels for unlabeled data is not available, we directly use soft predictions from the current model as pseudo labels for mixing;
\item \revise{FixMatch+VAT~\citep{Wang_2021_MedIA_Deep} is a semi-supervised learning method combining FixMatch~\citep{sohn_2020_fixmatch} and VAT~\citep{Miyato_2018_PAMI_VAT}, which has achieved state-of-the-art performance in large-scale medical image classification tasks. }
\end{itemize}
For all methods, we adopted their official implementation\footnote{For VAT, MixMatch, FixMatch, we re-implemented them in PyTorch as the original code repositories are based on Tensorflow.} and trained the same network with the same training setup \revise{(e.g. using the same pre-trained models)} for fair comparison. 
Quantitative results in Table \ref{tab:SSL_comparison} and qualitative results in Fig.~\ref{fig:SSL_comparison} shows that the proposed approach achieves the highest performance on the two segmentation tasks. Surprisingly, one interesting finding from Table~\ref{tab:SSL_comparison} is that when we have the same number of labeled and unlabeled images (i.e. N=11, M=11) to train the prostate segmentation network, the proposed method even exceeds the upperbound performance, e.g., 0.7325 vs 0.7288 in terms of average Dice score. This may be due to the presence of noisy labels in the prostate dataset (see the top-right block in Fig.~\ref{fig:SSL_comparison} for reference), which can affect the learning in the fully supervised setting. As semi-supervised learning does not fully rely on manual labels on the training dataset, it can therefore be more robust against noisy labels.

\begin{figure*}[t]
    \centering
    \includegraphics[width=0.8\textwidth]{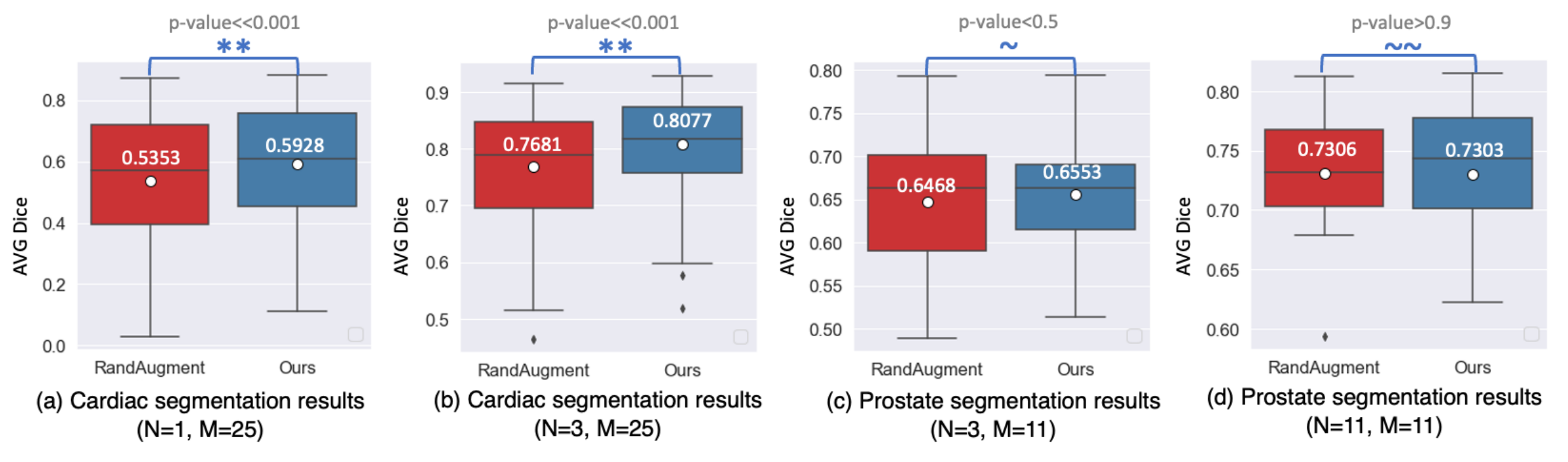}
    \caption{\revise{Boxplots of average Dice scores evaluated on the (a,b) cardiac and (c,d) prostate test sets using different composite data augmentation methods for consistency regularization: RandAugment~\citep{Cubuk_2020_RandAugment} and ours: \advchain. For ease of comparison, in \advchain, we limit the maximum number of selected transformations to 2, making it aligned with the recommended set-up in RandAugment. We trained the networks using different numbers of labeled (N) and unlabeled images (M). Compared to RandAugment, our proposed method \advchain\ achieves higher average Dice scores (see white numbers) in most cases even with a smaller set of transformations, especially when labeled data is extremely limited.}}
    \label{fig:rand_augmet_vs_ours_boxplots}
\end{figure*}

%% compare ours with rand augment
\revise{We further compared our data augmentation with the state-of-the-art random composite data augmentation method: RandAugment~\citep{Cubuk_2020_RandAugment} adopted in FixMatch~\citep{sohn_2020_fixmatch}, which employs a wide collection of image transformations including color inversion, translation, contrast adjustment~\footnote{We adopted the implementation of RandAugment provided in the official PyTorch website with its recommended set-up: \url{https://pytorch.org/vision/stable/generated/torchvision.transforms.RandAugment.html}.}. As shown in Fig.~\ref{fig:rand_augmet_vs_ours_boxplots} when we replaced our proposed data augmentation with RandAugment in our consistency-regularized method, the segmentation performance declines with lower average Dice scores on the cardiac segmentation task. The segmentation performance on the prostate segmentation tasks does not significantly outperform ours although RandAugment employs a larger number of image processing functions (\textit{autoContrast, equalize, solarize, color, posterize, contrast, brightness, sharpness, rotation, translation and shearing}~\citep{Cubuk_2020_RandAugment})}. 

\revise{We found that compared to \advchain, RandAugment focuses more on modifying the style of images. The geometric variations are quite limited compared to ours. In fact, RandAugment only considers basic spatial augmentation operations (e.g., rotation, translation) without applying any local deformations~\citep{Cubuk_2020_RandAugment}. Such a limitation is also shared in the other semi-supervised learning frameworks compared in our study, such as TCSM~\citep{Li_2020_TNNLS_transformation}. By contrast, our proposed method supports to generate diffeomorphic transformations to account for realistic morphological variations. The generated transformations are reversible, which allows to measure the prediction inconsistency in the original image space for ease of optimization. With adversarial training, \advchain\ observes the image content and takes the segmentation model's prediction to identify and deform the local structures of interest with increased variations (see Fig.~\ref{fig:adv_chain_vs_rand_chain}), which helps the segmentation model to better generalize across different populations with varied morphology. Of note, different from \advchain, RandAugment does not support adversarial training to optimize the transformation parameters as the underlying image transformation functions in Python Image Library (PIL)~\footnote{Python Image Library:~\url{https://pillow.readthedocs.io/en/stable/.}} do not support automatic differentiation. 
}

\begin{figure}[!htbp]
    \centering
    \includegraphics[width=\columnwidth]{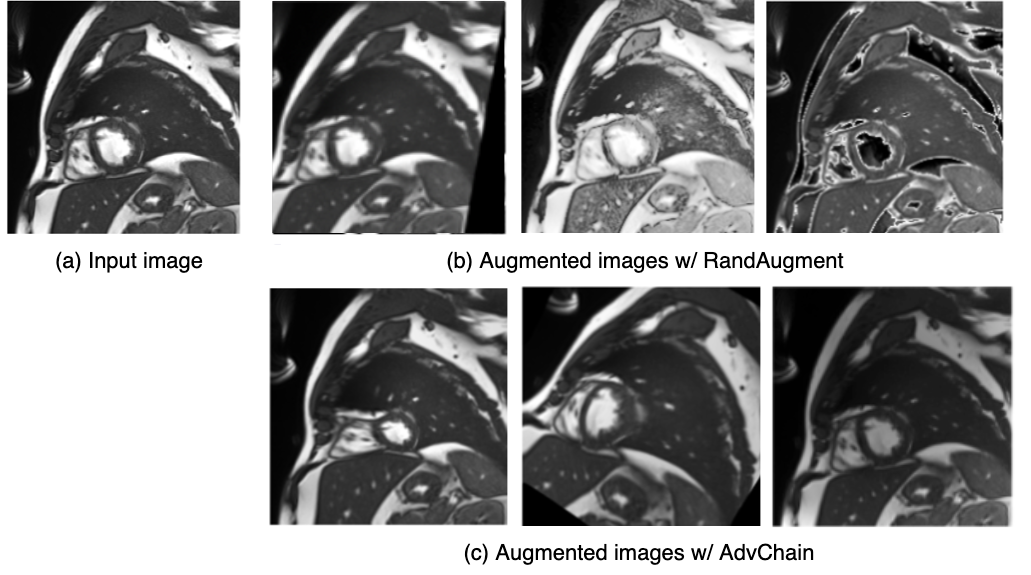}
    \caption{
    \revise{An input image (a) and (b) augmented images generated using RandAugment~\citep{Cubuk_2020_RandAugment} and (c) augmented images using our proposed adversarial data augmentation \advchain.}}
    \label{fig:rand_augmet_vs_ours_images}
\end{figure}

\revise{We attribute the efficacy of our method against other baseline methods mainly to 1) the increased data diversity and effectiveness with adversarial training applied to dynamic image transformations, see Sec.{\ref{abl:individual vs composite}}, \ref{abl:length and orders} and 2) the enhanced consistency regularization loss with a contour-based loss term, see Sec.\ref{abl:consistency loss}.}

\subsection{Ablation studies}
\subsubsection{Effects of adversarial training w/ individual augmentation and dynamically chained composite data augmentation}
\label{abl:individual vs composite}
To understand the effects of individual data augmentations
and the importance of the chain strategy for improved data diversity used in \advchain, we investigate the performance of our framework when applying
augmentations individually with a fixed type of transformation or with dynamic augmentation chains. Specifically, we trained the same network with each of the four transformations separately in the semi-supervised setting, and compared the results to their random counterparts (i.e. without adversarial training). Table~\ref{tab:random vs adv} shows the obtained results.  

%\begin{table}[!htbp]
%\centering
%\caption{Random vs adversarial data augmentation for consistency regularization. Experiments were performed on cardiac segmentation and prostate segmentation datasets in the supervised setting. n: number of training subjects. Specifically, we used three labeled cardiac subjects and eleven labeled prostate subjects for training, as the task complexities are different. Reported values are average Dice scores for foreground objects over multiple runs (5 runs for cardiac, three runs for prostate). }
%\label{tab:random vs adv}
%\begin{adjustbox}{width=0.9\columnwidth}
%\begin{threeparttable}
%\begin{tabular}{@{}ccccccc@{}}
%\toprule
% & Adversarial & Noise & Bias & Affine & Morph  & Compose (all)\\ \midrule
%\multirow{2}{*}{Cardiac (n=3)} & \crossmark & 0.8478
%&  0.8646
%&  0.8632
%&  0.8409& 
%0.8628\\
% & \checkmark 
% & \textbf{0.8508}
% & \textbf{0.8741} &  \textbf{0.8646}& \textbf{0.8543} & \textbf{0.8751} \\ \midrule
%\multirow{2}{*}{Prostate  (n=11)} & \crossmark & \textbf{0.8578} & 0.8481 & 0.8569 & 0.8598 & 0.8546 \\
% & \checkmark & 0.8524 & \textbf{0.8530} &  \textbf{0.8573} &\textbf{ 0.8669} & \textbf{0.8697}\\ \bottomrule
%\end{tabular}%
%\end{threeparttable}
%\end{adjustbox}
%\end{table}
%
%

\begin{table}[!ht]
\caption{\textbf{Random vs adversarial data augmentation with individual image transformations and chained transformations for consistency regularization.} Experiments were performed on cardiac segmentation and prostate segmentation datasets in the semi-supervised setting. For both tasks, we use only three labeled subjects. \revise{Reported values are mean (std) of average Dice scores across foreground classes over multiple runs (5 runs for cardiac, 3 runs for prostate)}.}
\label{tab:random vs adv}
\centering
\resizebox{0.9\columnwidth}{!}{%
\begin{tabular}{@{}cccccc@{}}
\toprule
\multicolumn{6}{c}{Cardiac} \\ \midrule
\multicolumn{1}{c|}{adversarial training} & {noise} & {bias} & {affine} & \multicolumn{1}{c|}{{morph}} & chain \\ \midrule
\multicolumn{1}{c|}{\crossmark} & 0.7706 (0.1066) & 0.7857 (0.0994) & 0.7704 (0.1123) & \multicolumn{1}{c|}{0.7836 (0.0940)	} & 0.7802	(0.1035) \\
\multicolumn{1}{c|}{\checkmark} & 0.7864 (0.0976) & 0.7955 (0.1013) & 0.7885 (0.1033) & \multicolumn{1}{c|}{0.8014 (0.0861)} & \textbf{0.8083 (0.0849)} \\ \midrule
\multicolumn{6}{c}{Prostate} \\ \midrule
\multicolumn{1}{c|}{adversarial training} & \multicolumn{1}{c}{{noise}} & \multicolumn{1}{c}{{bias}} & \multicolumn{1}{c}{{affine}} & \multicolumn{1}{c|}{{morph}} & \multicolumn{1}{c}{chain} \\ \midrule
\multicolumn{1}{c|}{\crossmark} & 0.5880 (0.1016) & 0.6093 (0.0817) & 0.6268 (0.0912) & \multicolumn{1}{c|}{0.6104 (0.0813)} & 0.6270 (0.0872)\\
\multicolumn{1}{c|}{\checkmark} & \multicolumn{1}{c}{0.6211 (0.0951)} & 0.6123 (0.0811) & 0.6294 (0.0897) & \multicolumn{1}{c|}{0.6408	(0.0784)} & \textbf{0.6492	(0.0789)} \\ \bottomrule
\end{tabular}%
}
\end{table}
In Table~\ref{tab:random vs adv}, we observe that individual data augmentations with adversarial training consistently outperform those corresponding ones without adversarial training, see row 1 vs. row 2, row 3 vs. row 4. Another finding is that adversarial training with morphological transformations always outperforms the other three individual data augmentations, highlighting the importance of introducing local anatomical variations to enhance the data variety. 

In both cardiac and prostate segmentation tasks, the proposed composite adversarial augmentation (\advchain) achieves the highest Dice scores on both tasks.  By contrast, the random-based composite data augmentation: \emph{chain} w/o adversarial training does not always outperform other random individual data augmentations. For example, on the cardiac segmentation performance, the average Dice score slightly drops from 0.7857 to 0.7802, compared to the one with random bias field augmentation.
{This highlights the benefits of applying adversarial training to optimizing dynamically chained transformations, which increases both the \emph{diversity} and \emph{effectiveness} of augmented data points to improve the network generalization for the downstream tasks.}

% \subsubsection{Impact of adversarial training}

\subsubsection{The generality of \advchain\ with different chained transformations of varied lengths and different orders}
\label{abl:length and orders}

\begin{figure}[!ht]
 \centering
    \includegraphics[width=0.95\columnwidth]{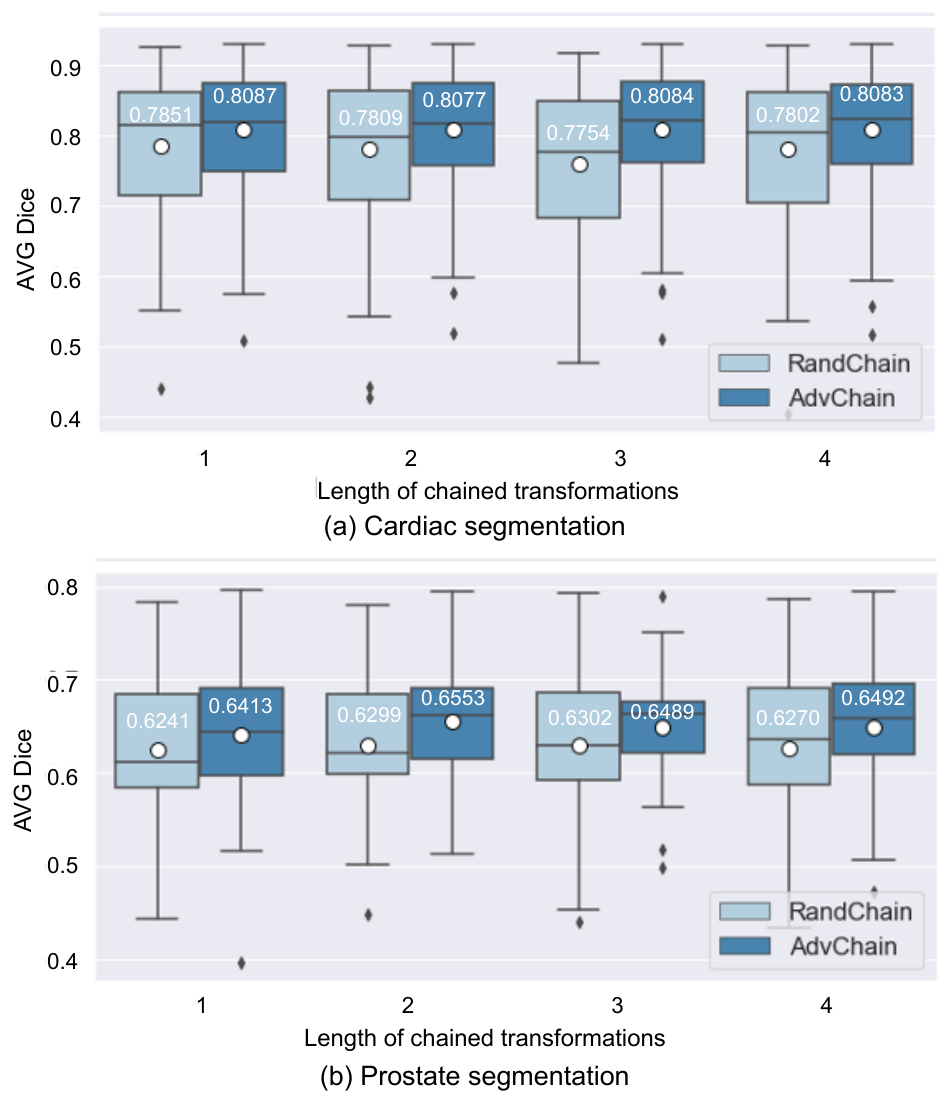}
    \caption{\revise{Boxplots of average Dice scores between the results of \randchain\ and \advchain\ for cardiac segmentation (a) and prostate segmentation (b). \advchain\ provides consistent improvements with chained transformations of different maximum lengths (1-4). White circles with numbers show the mean value of Dice scores across the test tests. Best viewed in color.}}
    \label{fig:different_lengths}
\end{figure}
\revise{
To verify the generality of \advchain\ with different types of chained transformations, we apply \advchain\ to optimizing chained transformations of different lengths, i.e. the maximum number of sampled transformations in a chain are fixed to a certain number (1/2/3/4) during the course of training in each experiment. Fig.~\ref{fig:different_lengths} plots the results on the cardiac test set using the same data setting (N=3, M=25) and prostate test set using the similar setting (N=3, M=11), respectively. We also plot the results with its downgraded variant (\randchain), i.e., \advchain\ without applying the adversarial optimization on the transformation parameters for comparison. Results show that \advchain\ consistently provides segmentation performance improvements regardless the change of chained lengths.}

\begin{figure}[t]
 \centering
    \includegraphics[width=0.95\columnwidth]{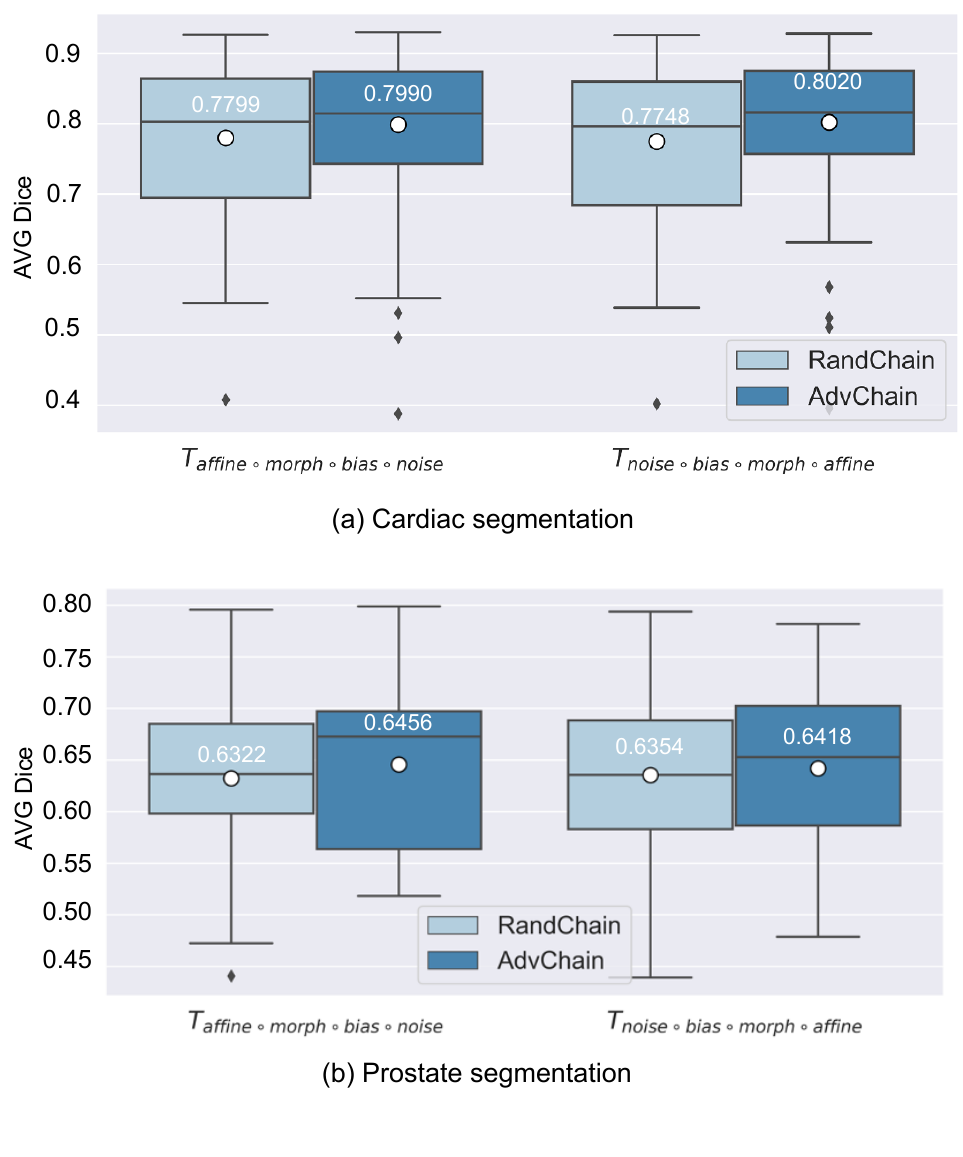}
    \caption{\revise{Boxplots of average Dice scores between the results of \randchain\ and \advchain\ with transformations chained in different orders. Compared to \randchain, \advchain\ boosts the segmentation performance with improved mean Dice scores and reduced outliers. White circles with numbers show the mean value of Dice scores across the test tests. Best viewed in color.}}
    \label{fig:different_orders}
\end{figure}
\revise{We also applied \advchain\ to optimizing the same set of transformations but chained in different orders. Since there are 24 different arrangements with the four transformations, we selected the two most common ones as a proof of concept: a) $T_{{affine} \circ {morph} \circ {bias} \circ {noise}}$: first apply photo-metric transformations  and then geometric transformations from local perturbations to global perturbations, similar to \citet{Chen_2019_SASHIMI_Intelligent, Zhao_2019_CVPR_oneshotDA}; b) $T_{{noise} \circ {bias} \circ {morph} \circ {affine}}$: the one chained in the opposite direction. Results shown in Fig.~\ref{fig:different_orders} confirm that \advchain\ provides consistent improvements with composite transformations chained in different orders, indicating the generality of \advchain\ with composite transformations chained in random orders for improved segmentation model performance.}

\begin{figure*}[t]
    \centering
    \includegraphics[width=\textwidth]{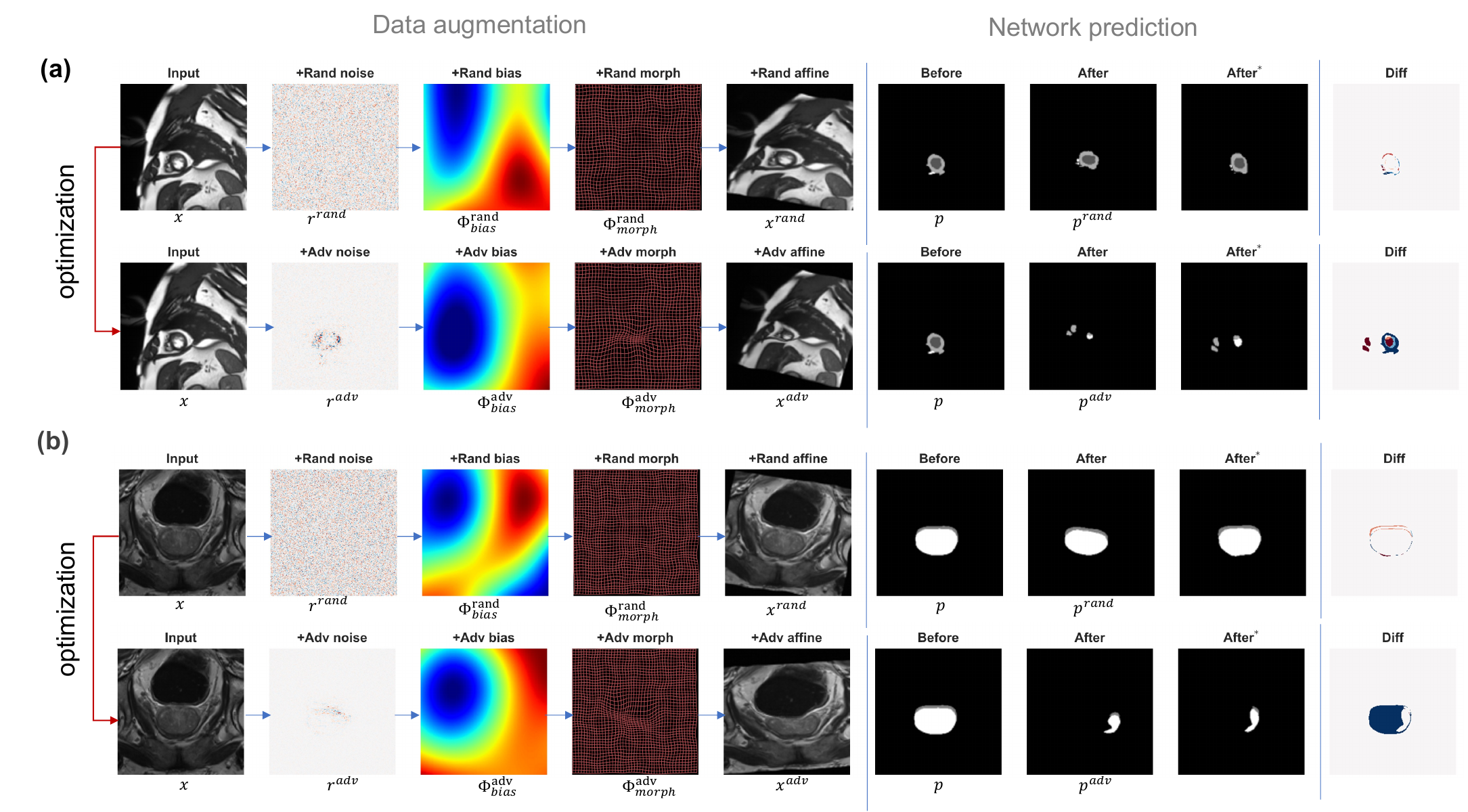}
    \caption{\revise{Optimizing a chain of data transformation parameters produces effective adversarial samples, which greatly alter network's predictions for (a) the cardiac segmentation task and (b) the prostate segmentation task, respectively. Here, the order of chained transformations is: $\augment_{\rm noise}\rightarrow \augment_{\rm  bias} \rightarrow \augment_{\rm  morph}\rightarrow \augment_{\rm  affine}$ ($\augment_{\rm  affine \circ \rm  morph \circ \rm  bias \circ \rm noise}$). Before/After: predictions before/after data augmentation. After$^*$: perturbed predictions which have been transformed back to the original image coordinates for consistency measurement. Best viewed in color and zoom in.}}
    \label{fig:adv_chain_vs_rand_chain}
\end{figure*}

Fig.~\ref{fig:adv_chain_vs_rand_chain} visualizes the data augmentation optimization process with the chain $T_{{affine} \circ {morph} \circ {bias} \circ {noise}}$ and network predictions before and after augmentations. From Fig.~\ref{fig:adv_chain_vs_rand_chain}, we can clearly see that after applying adversarial optimization, the optimized data augmentations (see the bottom row in each block) are more effective at perturbing network predictions compared to those with random initialization (the top row in each block). This is because adversarial data augmentation takes both model information and image content into account to augment images, which produces more informative, challenging samples to regularize the network. \revise{It is particularly evident when we compare adversarial noise ($\noise^{adv}$) and adversarial deformation ($\Phi_{\rm morph}^{adv}$) to their random initialized counterparts ($\noise^{rand}, \Phi_{\rm morph}^{rand}$).  We can see that adversarial data augmentation can identify and focus more on attacking/deforming local target structures in images to fool the network to make inconsistent predictions}. Augmenting images with these adversarial transformations contribute to stronger consistency regularization to enforce the network to be invariant under photometric transformations and equivariant under geometric transformations.

\revise{
In Fig~\ref{fig:different_orders}, it is interesting to notice that \advchain\ with $T_{{noise} \circ {bias} \circ {morph} \circ {affine}}$ yields slightly better performance compared to the one with $T_{{affine} \circ {morph} \circ {bias} \circ {noise}}$ on the cardiac segmentation task. For the prostate segmentation, \advchain\ w/ $T_{{affine} \circ {morph} \circ {bias}}$ achieves higher segmentation. Similarity, \advchain\ achieves slightly better performance when the maximum chained length is fixed to 1 for cardiac segmentation and 2 for prostate segmentation, respectively, as shown in Fig.~\ref{fig:different_lengths}. We also found that the optimum maximum chain length depends on not only the task but also the selection of training set, see Fig.~\ref{fig:optimum_chain_length_across_dataset_selections} in the appendix. It is possible that better performance can be further achieved by identifying the optimum maximum length and the optimum arrangement (taking the validation set performance into account) to improve segmentation performance for a \emph{specific} task. Yet since the search space
can be extremely large and the policy optimization requires extraordinary high computational costs~\citep{Cubuk_2018_AutoAugment}, we randomly generate arbitrarily chained transformations to explore all different kinds of possibilities as a trade-off between efficiency and effectiveness for general segmentation tasks. 
}

\subsubsection{Effect of different consistency loss functions}
\label{abl:consistency loss}

\begin{table}[!ht]
\caption{Effect of different distance functions for consistency loss function $\RegLoss$. Reported scores are average Dice scores over segmented structures. We also report the mean and standard deviation of average Dice scores over all foreground classes (AVG) for each task. N: number of labeled subjects, M: number of unlabeled subjects.}
\label{tab:consistency_loss_functions}
\resizebox{\columnwidth}{!}{%
\begin{tabular}{l|cccc|ccc}
\hline
 & \multicolumn{4}{c|}{Cardiac (N=3, M=25)} & \multicolumn{3}{c}{Prostate (N=3, M=11)} \\ \hline
Consistency loss functions & LV & MYO & \multicolumn{1}{c|}{RV} & \multicolumn{1}{l|}{AVG} & PZ & \multicolumn{1}{r|}{CZ} & AVG \\ \hline
$\RegLoss_{\Distance_{KL}}$ & 0.8635 & 0.8429 & \multicolumn{1}{c|}{0.6747} & 0.7937 & 0.5053 & \multicolumn{1}{r|}{0.7663} & 0.6358 \\
$\RegLoss_{\Distance_{KL+Contour}}$ & 0.8655 & 0.8445 & \multicolumn{1}{c|}{0.6744} & 0.7948 & 0.4989 & \multicolumn{1}{r|}{0.7729} & 0.6359 \\
$\RegLoss_{\Distance_{MSE}}$ & 0.8660 & 0.8450 & \multicolumn{1}{c|}{0.7043} & 0.8051 & 0.5156 & \multicolumn{1}{r|}{\textbf{0.7744}} & 0.6450 \\
$\RegLoss_{\Distance_{MSE+Contour}}$ (proposed) & \textbf{0.8708} & \textbf{0.8469} & \multicolumn{1}{c|}{\textbf{0.7072}} & \textbf{0.8083} & \textbf{0.5243} & \multicolumn{1}{r|}{0.7742} & \textbf{0.6492} \\ \hline
\end{tabular}
}
\end{table}
We further compared the proposed with the other three different distance functions to highlight the superiority of the proposed inconsistency regularization $\RegLoss$. The three different distance functions have been commonly used in the literature for semi-supervised learning, which are:
\begin{itemize}
    \item $\RegLoss_{\Distance_{KL}}$, the regularization loss used in VAT~\citep{Miyato_2018_PAMI_VAT} and FixMatch~\citep{sohn_2020_fixmatch}, FixMatch+VAT~\citep{Wang_2021_MedIA_Deep}, where $\Distance_{KL}$ is Kullback–Leibler divergence (KL) loss: $ \RegLoss_{\Distance_{KL}}=\Distance_{KL}(\p,\p') = 1/n\sum_{i=1}^{n}\sum_{c=1}^C{\p(i)^{(c)} \log \frac{\p(i)^{(c)}}{\p'(i)^{(c)}}}$ where $n$ is the number of pixels in the image;
    \item $\RegLoss_{\Distance_{MSE}}$, the regularization loss used in TCSM~\citep{Li_2020_TNNLS_transformation} and  MixMatch~\citep{berthelot_2019_mixmatch}, where $\Distance_{MSE}$ is the mean squared loss;
    \item  $\RegLoss_{\Distance_{KL+Contour}}$, where $\Distance_{KL+Contour}$ consists of the $\Distance_{KL}$ loss and the contour-based loss $\Distance_{Contour}$ used in our previous work~\citep{chen_2020_realistic}.
\end{itemize}
And the proposed one is denoted as $\RegLoss_{\Distance_{MSE+Contour}}$ for clarity. We ran experiments on the two tasks. Results are shown in Table~\ref{tab:consistency_loss_functions}. Compared to the other three loss functions, the proposed one $\RegLoss_{\Distance_{MSE+Contour}}$ outperforms the other three in most cases, and achieves the highest average Dice scores in both segmentation tasks. Adding contour-based loss ($\RegLoss_{\Distance_{KL+Contour}}$,$\RegLoss_{\Distance_{MSE+Contour}}$) in general provides better performance than their corresponding standalone counterpart (i.e. $\RegLoss_{\Distance_{MSE}},
\RegLoss_{\Distance_{KL}}$), highlighting the benefits of taking additional boundary information for consistency regularization.

\subsubsection{Effect of number of labeled images}
\begin{figure}[t]
    \centering
    \includegraphics[width=0.85\columnwidth]{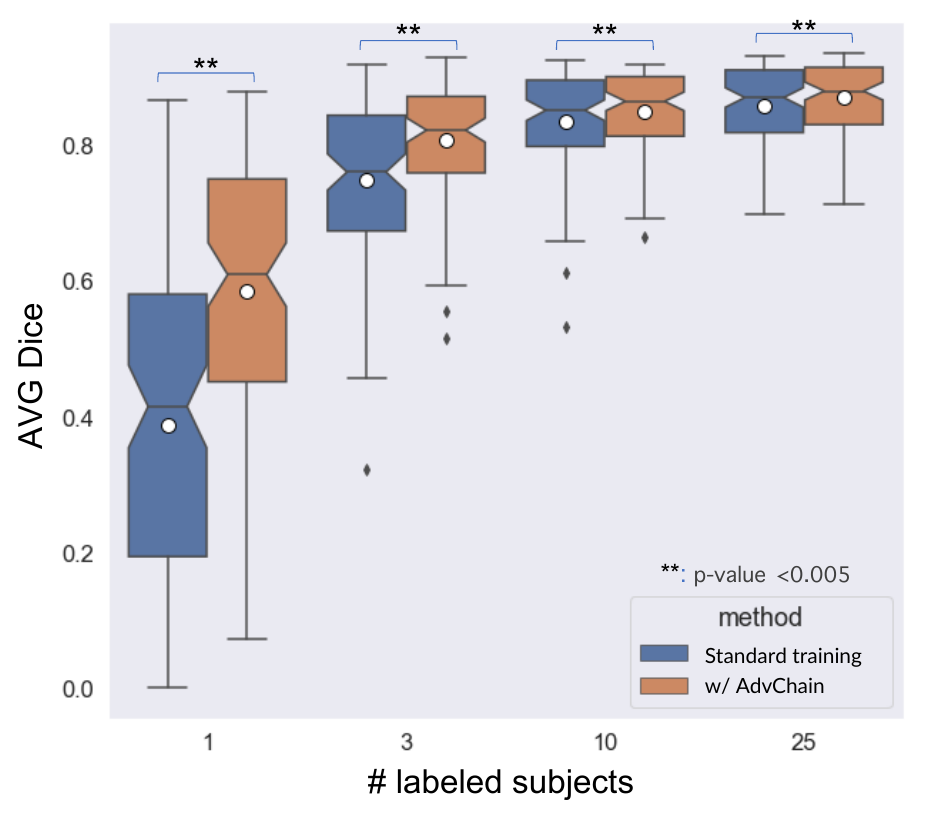}
    \caption{Semi-supervised learning results on the cardiac test set with networks trained using different number of labeled subjects and the same unlabeled set (25 subjects). Standard training: supervised training using only labeled images. Best viewed in color. %Numbers shown on the boxplots are the mean values of Dice scores for network predictions using different methods.
    % The upperbound mean Dice score (training the same network using 25+25 labelled images from both labelled and unlabelled set) is 0.8938.
    }
    \label{fig:cardiac_varing_subjects}
\end{figure}

In Figure~\ref{fig:cardiac_varing_subjects}, we report segmentation performance of our methods under different data settings on the cardiac segmentation task. Networks were trained with a different number of labeled subjects and the same unlabeled dataset ($M=25$). As expected, the performance of semi-supervised learning improves when more labeled training images are included. Compared to standard training (trained without consistency regularization), the proposed method (\advchain) consistently provides significant performance improvement
across all settings. The performance gain is particularly evident in the extremely one-shot setting (N=1). \revise{The performance gap between the standard
training and ours narrows when adding more labeled
images into training, which is consistent with the finding reported in other semi-supervised learning frameworks~\citep{Li_2020_TNNLS_transformation}.}

\begin{table*}[t]
\caption{Results of low-shot supervised learning on the cardiac and prostate datasets. Reported values are average Dice scores.}
\label{tab:supervised}
\centering
\resizebox{0.78\textwidth}{!}{%
\begin{tabular}{@{}c|ccc|ccc|cc|cc@{}}
\toprule
 & \multicolumn{6}{c|}{Cardiac} & \multicolumn{4}{c}{Prostate} \\
 & \multicolumn{3}{c|}{N=1, M=0} & \multicolumn{3}{c|}{N=3, M=0} & \multicolumn{2}{c|}{N=3, M=0} & \multicolumn{2}{c}{N=11, M=0} \\ \midrule
 & LV & MYO & RV & LV & MYO & RV & PZ & {CZ} & PZ & CZ \\
 Standard training & 0.5155 & 0.4290 & 0.2201 & 0.8269 & 0.7905 & 0.6288 & 0.3017 & 0.6278 & 0.5057 & 0.8091 \\
% +MixMatch~\citep{berthelot_2019_mixmatch} &  0.5793 & 0.4703 & 0.2463  &  0.8436 & 0.8149 & 0.6105  & 0.3272 & 0.6820 & 0.5600 & 0.8322 \\
% +TCSM~\citep{berthelot_2019_mixmatch} &  - & - & -  &  -   & - & -  & - & 0.- & 0.- & 0.- \\
% +VAT~\citep{Miyato_2018_PAMI_VAT} & 0.5709 & 0.4627 & 0.2852 & \textbf{0.8438} & 0.8080  & 0.6182  & 0.3178 & 0.6583 & 0.5233 & 0.8223\\  
\randchain & 0.5581 & 0.4570 & 0.2699 & 0.8183 & 0.7857 & 0.6123 & 0.3962 & {0.7570} & 0.5641 & 0.8385 \\
\advchain & \textbf{0.6093} & \textbf{0.5022} & \textbf{0.3079 } &\textbf{0.8435} & \textbf{0.8122} & \textbf{0.6473} & \textbf{0.4192} & {\textbf{ 0.7600}} & \textbf{0.5720} & \textbf{0.8450} \\ \bottomrule
\end{tabular}%
}
\end{table*}
\subsection{Supervised learning with extremely low data settings}
We evaluate the performance of the proposed method under extremely low data settings, where there is no unlabeled data available ($M=0$). Results are shown in Table~\ref{tab:supervised}. It is clearly that the proposed \advchain\ consistently outperforms the competitive baseline method \randchain\  on the two tasks by a large margin. The results confirm that in the scenario where training data is limited, the proposed method still enhances model training significantly. This indicates the great potential of the proposed method to alleviate data scarcity problem.

% \subsubsection{\advchain\ improves cross-domain robustness}
% We further trained the segmentation network using a large scale training set (3,975 subjects) from the UK Biobank~\footnote{t http://imaging.ukbiobank.ac.uk/.} and tested it on a test set of 600 subjects from the same study. We compared \advchain\ with the current state-of-the-art method~\citep{Chen_2020_Frontiers} using the same configurations. its downgraded version \randchain and 

\section{Discussion}
In this work, we have presented a novel adversarial data augmentation method, which is capable of introducing both realistic photometric and geometric transformations to improve the generalization capability for neural network-based medical image segmentation of MR images. The proposed method enhances several aspects of previous data augmentation and regularization schemes. \revise{Compared to VAT~\citep{Miyato_2018_PAMI_VAT}, RandAugment~\citep{Cubuk_2020_RandAugment} and the data augmentation in TCSM \citep{Li_2020_TNNLS_transformation}, the proposed method provides counterpart samples with more realistic variations in medical imaging, including challenging local intensity variations (bias fields) and morphological changes (diffeomorphic deformations).} Compared to data-mixing based methods such as Mixmatch~\citep{berthelot_2019_mixmatch}, which generates unrealistic mixed images with linear interpolation to ensure the `linearity' of the network, the proposed method applies physics-based transformation models to generate a diverse set of extrapolated data points around each input, which can be viewed as a way to encourage the `local smoothness' under various local perturbations. We believe that local smoothness is a better regularization for segmentation tasks, as it encourages the network to incorporate human perception, clustering perceptually similar images for decision making. In particular, it \emph{strengthens the network's invariance against photometric transformations and equivariance under geometric transformations}, where the two properties are highly desirable for model generalization. On the segmentation tasks for cardiac and prostate MR images, we demonstrated that the proposed method has great potential to reduce the annotation effort, outperforming competitive baseline methods in both low-shot supervised settings and semi-supervised settings. 

%% comparison between ours and GAN-based
We notice that there are concurrent  works using GAN~\citep{Chaitanya_2019_IPMI} and adversarial training~\citep{Gao_2021_IPMI_Enabling_Data_Diversity} to find effective  photometric and geometric transformations for data augmentation. These methods cannot be directly compared to ours as they require  training additional neural networks. Since GANs are essentially large neural networks, they still require a large number of training images to avoid over-fitting. \revise{And their methods can not be applied to optimize dynamically chained transformations as it suffers from the training instability problem with a set of randomly stacked GANs.} Our method, by contrast, is more \emph{flexible} and \emph{data efficient}, as it only employs a small set of explainable and controllable parameters and can be used even in extremely low data settings (e.g., only 1 or 3 labeled subjects, no access to unlabeled subjects). Also, training GAN requires considerable computational resources and expertise to tune hyperparameters and can be very unstable~\citep{Gulrajani_2017_NIPS_Improved}. The proposed adversarial data augmention by contrast, can be directly used as a plug-in lightweight module to support training segmentation pipelines.

%% limitation and future direction
\textbf{Limitations:} One limitation of the proposed method is that it still requires expertise to explicitly specify the magnitude constraints for the employed parameters to ensure the naturalism of augmented images. Yet, how to automatically find optimal data augmentation policy (data augmentation operations and associated probabilities, magnitudes, \revise{the order of chained transformations, the optimum chain length}) is still an active research area~\citep{Cubuk_2018_AutoAugment,Zhang_2020_ICLR_Adversarial_AutoAugment,Shorten_2019_Big_Data_Survey}. In recent years, there has been an emerging research topic focusing on automatic data augmentation (Auto DA), which in general requires an external RNN controller to find optimal probabilities and magnitudes for a group of image transformations for a particular dataset~\citep{Cubuk_2018_AutoAugment,Shorten_2019_Big_Data_Survey}. Combining Auto DA with the proposed method may further automate the process with higher accuracy.

\revise{To further enhance the effectiveness of the proposed method for medical image segmentation tasks, one can also consider employing more advanced segmentation network architectures to increase its representation learning capacity for improved segmentation accuracy and integrating \advchain\ with other advanced techniques to solve potential additional challenges.} For example, medical image segmentation often suffers from marked class-imbalance (long-tail problem), which may skew the performance of the segmentation model. In this work, we applied weighted supervised loss, as a common practice. It is worthwhile to explore more advanced class-imbalance invariant techniques, e.g., suppressed consistency loss~\citep{Hyun_2020_ICML} for further improvements. On the other hand, medical images are typically grayscale images with poor image contrast where the anatomical structures may have very blurry contours. This increases the burden of producing reliable predictions (pseudo labels) on unlabeled images before applying perturbations for consistency regularization. To improve the reliability of pseudo labels, one can adopt an iterative training procedure, which distills previously learned knowledge into a neural network with equal or larger capacity to boost model performance on label estimation~\citep{Xie_2019_Arxiv_Noisy_Student,Zoph_2020_Arxiv_Rethinking_Pretraining}. Also, it is interesting to introduce a pseudo label assessment module to select high quality pseudo labels for more effective uncertainty-aware consistency regularization~\citep{Xia_2020_CV,Xia_2020_MedIA_Uncertainty,Liu_2021_Certainty,Wang_2021_MedIA_Self_paced,Yu_2019_MICCAI}. We will explore these
extensions in future work.

\section{Conclusion}
\label{SEC: conclusion}
This work tackles the challenging task of multi-class segmentation on MR images, given very limited number of labeled subjects. \revise{A novel adversarial data augmentation method has been presented, which jointly optimizes a \emph{dynamic data augmentation} module and the segmentation network to better leverage labeled and unlabeled data for improved model generalization. The proposed data augmentation method is capable of improving both data effectiveness and diversity with challenging complex data variations based on photo-metric and geometric transformations (Sec.~\ref{abl:individual vs composite}, Sec.~\ref{abl:length and orders}), simulating realistic image appearance and anatomical variations that could exist in MR imaging. Our work also highlights the importance of 1) introducing adversarial diffeomorphic deformations for improved data diversity and effectiveness (Sec.~\ref{abl:individual vs composite}), which has not been explored in prior consistency regularization-based methods. We have also demonstrated the effectiveness of adding the contour-based consistency loss for more comprehensive inconsistency measurement to inform network training, see Sec.~\ref{abl:consistency loss}}. 

The whole framework can be used as a plug-in module to facilitate supervised and semi-supervised learning and is generic for
MR image segmentation tasks. \revise{With only four types of photo-metric and geometric transformations, we have demonstrated
its great data efficiency on two different tasks in challenging low-shot semi-supervised settings, outperforming several strong consistency-regularized methods in different scenarios. The proposed method even outperforms the state-of-the-art composite data augmentation method (RandAugment~\citep{Cubuk_2020_RandAugment}) in most cases. The flexibility and the generic nature of \secondrevise{\advchain}\ opens the door to incorporate more image transformations to better reflect the imaging variations in the real world and thus \advchain\ has the potential to be applied to different imaging modalities and different data-driven medical imaging applications, such as image registration~\citep{uzunova_2017_MICCAI_training} and image reconstruction~\citep{cheng_2020_MIDL_addressing_FP_in_recon}. We leave that for future work.
}

\section*{Acknowledgment}
 This work was supported by two EPSRC Grants (EP/P001009/1, EP/R005982/1) and the ERC Grant (884622). W. Bai was supported by EPSRC DeepGeM Grant (EP/W01842X/1).

\bibliographystyle{model2-names.bst}\biboptions{authoryear}
\bibliography{mybib}
\section*{Appendix}
\renewcommand{\thefigure}{A\arabic{figure}}
\setcounter{figure}{0}
\begin{figure*}[hbt!]
 \centering
    \includegraphics[width=\textwidth]{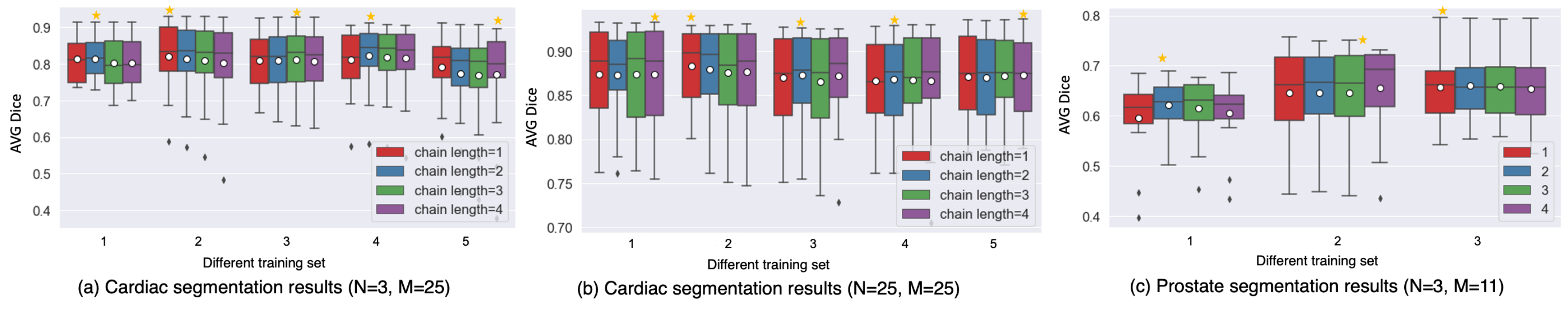}
    \caption{\textbf{Optimal maximum chain length of \advchain\ depends on the selection of training set, training set size and the segmentation task.} Here, we plot boxplots of average Dice scores with the segmentation network trained with \advchain\ using different selections of labeled set across different tasks. We varied the maximum chain lengths (1-4) to search for the optimum setting that achieves the highest average Dice scores on the test set. A yellow star in each group indicate the optimum chain length for a specific setting. We find that the optimum value varies across different tasks and different labeled set selections, which can be observed on the cardiac and prostate segmentation tasks, see (a), (c), respectively. This phenomenon still exists even when we increased the number of labeled subjects from 3 to 25 for the cardiac segmentation task (b). There is no consensus on the optimum maximum chain length across different selections of labeled sets. N: number of labeled images, M: number of unlabeled images.}
    \label{fig:optimum_chain_length_across_dataset_selections}
\end{figure*}
\clearpage

\end{document}